\documentclass{article}
\usepackage{graphicx}
\usepackage[margin=1in]{geometry}
\usepackage{tikz}
\usepackage{tenebraelux}
\usepackage{amsmath}
\usepackage{amsthm}
\usepackage{amssymb}
\usepackage{amsfonts}
\usepackage{mathtools}
\usepackage{tabularx}
\usepackage{booktabs}
\usepackage{multirow}
\usepackage{thm-restate}
\usepackage{hyperref} % load (almost) last
\hypersetup{colorlinks,linkcolor={blue},citecolor={blue},urlcolor={blue}}

\newtheorem{theorem}{Theorem}
\newtheorem{lemma}{Lemma}
\newtheorem{prop}{Proposition}

\newcommand{\problemStatement}[3]{%
  \begin{center}
  \begin{tabularx}{\columnwidth}{@{}lX@{}}
  \toprule
  \multicolumn{2}{@{}c@{}}{#1}\tabularnewline
  \midrule
  \bfseries Instance:    & #2 \\
  \bfseries Question: & #3 \\
  \bottomrule
  \end{tabularx}
  \end{center}
}

\newcounter{tlpanel}
\renewcommand\thetlpanel{(\alph{tlpanel})}

\newcommand\tlpanellabel{\par\smallskip\refstepcounter{tlpanel}\thetlpanel}
\newlength\tleqcell
\newsavebox\tleqbox
\newcommand\tleqpair[3]{%
  \sbox\tleqbox{%
    \begin{tikzpicture}[baseline=(current bounding box.center)]
      \tlgrid[cell=\tleqcell]{#1}
      #2
    \end{tikzpicture}%
    \;$\Rightarrow$\;%
    \tlboard[cell=\tleqcell]{#3}}%
  \ifdim\wd\tleqbox>\linewidth
    \resizebox{\linewidth}{!}{\usebox\tleqbox}%
  \else
    \usebox\tleqbox
  \fi}

\title{Cal's Favourite Game Is NP-Hard:\\On the Hardness of Tenebrae Lux's Sliding Puzzles}

\author{Sarosh Adenwalla \footnotemark[1] \and Ziad Ismaili Alaoui \thanks{University of Liverpool, United Kingdom. \texttt{\{sarosh.adenwalla, ziad.ismaili-alaoui\}@liverpool.ac.uk}}\;\;\thanks{Philipps-Universität Marburg, Germany.}}

\date{}

\begin{document}

\maketitle

\begin{abstract}
    \noindent We study the computational complexity of a generalised version of the minigame Tenebrae Lux, featured in Grand Theft Auto VI.

    An instance of \textsc{Tenebrae Lux} consists of an $h \times w$ board whose cells are either empty or contain a \emph{pyramid}, a \emph{stack}, or a \emph{flat}. Initially, no pyramid is orthogonally adjacent to a stack or a flat. A move selects a pyramid or stack and slides it in a cardinal direction through consecutive empty cells until it reaches the board boundary or stops immediately before another piece. Each move must change the selected piece's position. At its destination, a moved pyramid destroys all orthogonally adjacent stacks and flats, and is itself destroyed if it destroyed a stack. A moved stack destroys all orthogonally adjacent pyramids, and is itself destroyed if it destroyed a pyramid. Flats cannot move.

    \textsc{Tenebrae Lux} asks whether there exists a finite sequence of moves that empties all cells; we call such a sequence \emph{clearing}. We show that this problem is NP-hard, even on boards with exactly one pyramid and one stack, via a reduction from \textsc{Hamiltonian Cycle} on simple 2-out-regular directed graphs. Since solvable boards with a bounded number of movable pieces admit clearing sequences of polynomial length, this restriction is in fact NP-complete. In general, the problem lies in PSPACE, and it is solvable in polynomial time when the total number of pieces is bounded. We further show that there always exists a clearing sequence when there are no flats and the numbers of pyramids and stacks are equal; our proof is constructive and yields a clearing sequence of at most $k(w+2)$ moves in polynomial time, where $k$ is the number of pyramids.
\end{abstract}

\tableofcontents

\newpage

\section{Introduction}

Grand Theft Auto VI, developed and published by Rockstar Games and released in November 2026, is an action-adventure game. Among many different activities that can be done in its open world, such as buying flowers, decorating bedrooms, and blowing up half a city, players can use an in-game console to play several fictional video games. One of them is Tenebrae Lux, a first-person psychological puzzle adventure credited to the fictional developer Johannes Zuch, in which the player explores an island and solves various puzzles. Puzzles from popular games have long been a fruitful source of problems in computational complexity, from classic puzzles such as Rush Hour~\cite{flake2002rush} to the video games of Nintendo~\cite{aloupis2015nintendo} and match-three games such as Candy Crush~\cite{walsh2014candy,guala2014bejeweled}; we refer to the surveys of Demaine and Hearn~\cite{demaine2009playing}, Kendall, Parkes and Spoerer~\cite{kendall2008survey} and Uehara~\cite{uehara2023puzzles}, and to the book of Hearn and Demaine~\cite{hearn2009games}, for an overview (more on that in Subsection~\ref{ss:related}).

\begin{figure}[!ht]
    \centering
    \includegraphics[width=0.75\linewidth]{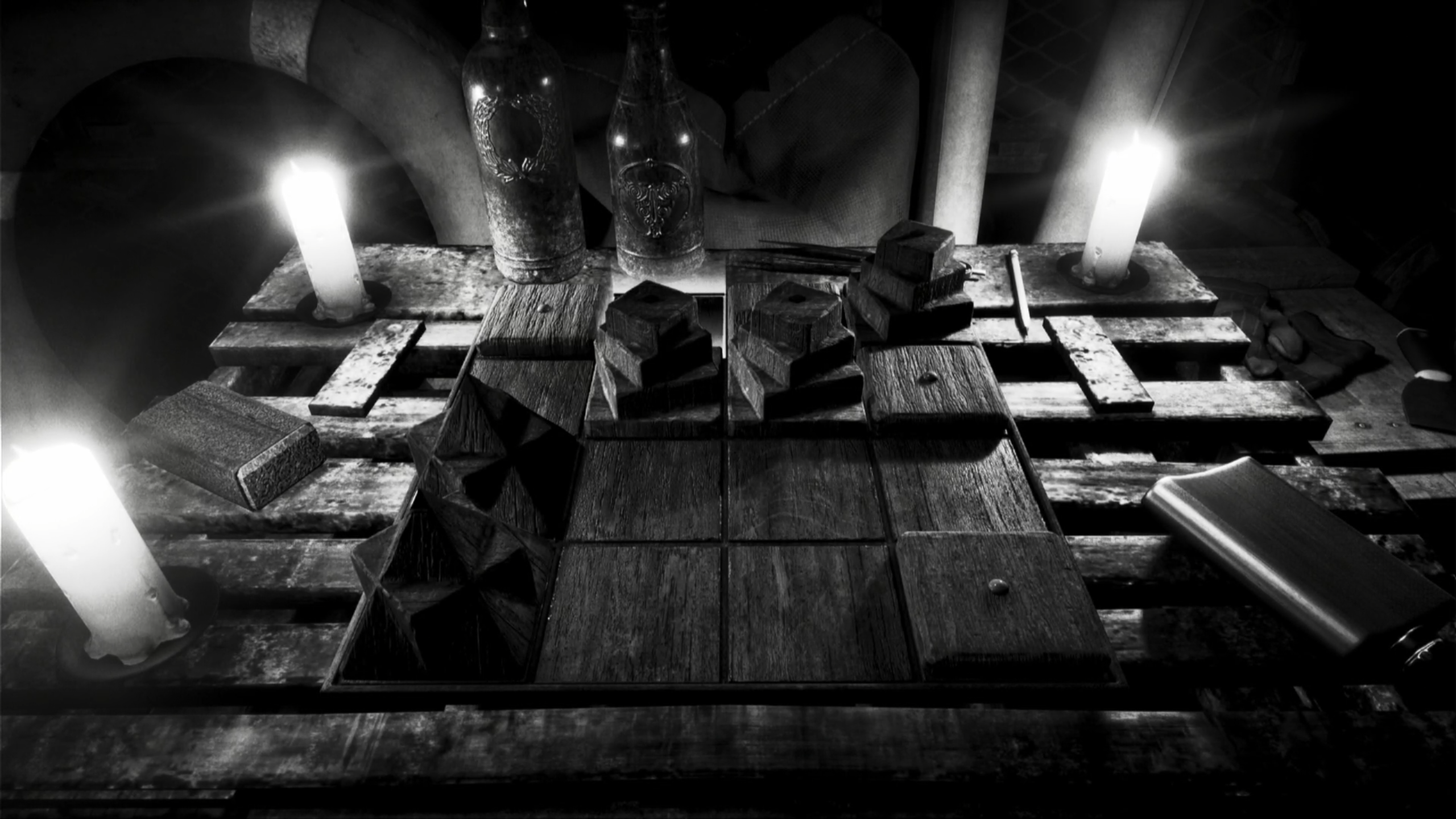}
    \caption{A screenshot of a sliding puzzle in Tenebrae Lux (note that this is originally a black-and-white game).}
    \label{fig:example-in-game}
\end{figure}

In this paper, we focus on the sliding puzzles encountered within Tenebrae Lux. These puzzles are played on a $4 \times 4$ board whose cells are either empty or occupied by one of three types of pieces: \emph{pyramids}, \emph{stacks}, and \emph{flats}. Each cell contains at most one piece, and two cells are adjacent if they share an edge (and thus cannot be adjacent diagonally). The player can only move stacks and pyramids. Figure~\ref{fig:example-in-game} shows an example of a board encountered in the game. A \emph{move} selects a pyramid or stack and slides it horizontally or vertically through consecutive empty cells. The selected piece must travel at least one cell and continues until it reaches either the board boundary or the cell immediately before another piece; thus, a piece cannot be stopped at an arbitrary intermediate cell. Once the piece comes to rest, every pyramid destroys all stacks and flats adjacent to it, and a pyramid that is adjacent to a stack is destroyed as well. In particular, a pyramid that removes only flats survives. All destructions depend only on adjacency at the destination; pieces passed during the slide are unaffected. Destroyed pieces leave empty cells that can be used in subsequent moves. The objective of the game is to destroy all pieces on the board. Subsection~\ref{ss:prelim} states these rules formally, and Figure~\ref{fig:ingame-solution} shows a shortest solution of an example board.

\noindent Since a $4 \times 4$ board has at most $4^{16}$ configurations, we can, in principle, decide in constant time whether an original puzzle can be solved, and how to solve it, using a precomputed lookup table. We therefore study a generalised version of the sliding puzzles in Tenebrae Lux that preserves the movement and destruction rules while allowing the board dimensions to be part of the input. We formalise the resulting decision problem as follows.

\problemStatement{\textsc{Tenebrae Lux}}
  {An $h \times w$ board whose cells are either empty or contain a stack, a pyramid, or a flat, such that initially no pyramid is adjacent to a stack or flat.}
  {Can all pieces be destroyed by a finite sequence of moves?}

\subsection{Our Results and Proof Overview}

It turns out that \textsc{Tenebrae Lux} is NP-hard, and remarkably, this remains true when the board contains a single pyramid and a single stack, every other piece being a flat.

\begin{restatable}{theorem}{main}
\label{thm:main}
\textsc{Tenebrae Lux} is NP-hard, even on boards containing exactly one pyramid and one stack.
\end{restatable}

\noindent When the number of movable pieces is bounded by a constant, a shortest solution has polynomial length, so the restriction in Theorem~\ref{thm:main} is in fact NP-complete. In general, \textsc{Tenebrae Lux} lies in PSPACE by Savitch's theorem~\cite{savitch1970relationships} (Proposition~\ref{prop:membership}), and we leave open whether it is PSPACE-complete. Now, to give a flavour of the reduction, we reduce from \textsc{Hamiltonian Cycle} on simple 2-out-regular directed graphs, in which every vertex has exactly two outgoing arcs. Rusu~\cite{rusu2022hamiltonian} showed this problem to be NP-complete. We first split every vertex $x$ into a copy $x^-$, which keeps the incoming arcs of $x$, and a copy $x^+$, which keeps its outgoing arcs, and we add the arc $x^- \to x^+$ for every vertex $x$. Fixing an arbitrary vertex $v_\rho$, the Hamiltonian cycles of the original graph then correspond exactly to the paths that start at $v_\rho^+$, visit every copy once, and end at $v_\rho^-$. The unique pyramid starts at $v_\rho^+$.

The board, whose size is polynomial in that of the graph, assigns three flats to each vertex: an arrival flat, a departure flat, and a turning flat between them (the starting vertex $v_\rho$ needs neither a departure flat nor a turning flat). Each arc receives one further turning flat. Recall that a pyramid destroys every flat adjacent to it as soon as it stops, and is itself not destroyed. Visiting a vertex is then equivalent to destroying its flats, and the turning flats are the only places where the pyramid can stop and change direction for doing so. Crucially, they can be used only \emph{once}: each vertex is assigned its own rows, and after the pyramid leaves the row of a vertex, it can never stop in that row again. It must therefore destroy both turning flats of the outgoing arcs of a vertex during its single visit, and the one destroyed last determines the next vertex. Destroying every flat is thus possible only by following arcs of the graph, visiting each vertex exactly once, and finally returning to $v_\rho$, which is exactly what a Hamiltonian cycle is.

A unique stack occupies the bottom-left corner, enclosed by two flats, and serves as the exit of the tour. The pyramid may technically reach it at any time, but contact destroys both pieces, so this clearly must be the final move; it empties the board only if every flat has already been destroyed. Consequently, the graph has a Hamiltonian cycle if and only if the board can be cleared.

As a positive result, we show that any board that does not contain flats and contains an equal number of pyramids and stacks must have a solution.

\begin{restatable}{theorem}{equalalwayspossible}
\label{thm:equal}
    \textsc{Tenebrae Lux} is always solvable when there are no flats and as many pyramids as stacks.
\end{restatable}

\noindent The proof is constructive and gives a solution in polynomial time.

\subsection{Related Work}
\label{ss:related}

Tenebrae Lux belongs to the family of puzzles in which pieces slide until they are blocked. In Atomix, the player slides atoms on a board with fixed walls to assemble a target molecule, and deciding whether a solution even exists is PSPACE-complete~\cite{holzer2004atomix}. Hartline and Libeskind-Hadas~\cite{hartline2003motion} showed that a generalisation of Lunar Lockout with fixed blocks is PSPACE-complete, and Balanza-Martinez et al.~\cite{balanza2024ricochet} gave a simple PSPACE-completeness proof for the relocation problem in Ricochet Robots. In all of these games, the pieces persist and the goal is to reach a target. In Tenebrae Lux, however, every interaction destroys pieces, so the number of pieces can only decrease. This makes progress irreversible, which is what allows a bounded number of movable pieces to place the problem in NP. Destruction is also central to match-three games (such as Candy Crush), whose complexity was studied by Walsh~\cite{walsh2014candy} and Gualà, Leucci and Natale~\cite{guala2014bejeweled}; there, however, the removed pieces are replaced by falling ones rather than leaving empty cells.

\subsection{Preliminaries}
\label{ss:prelim}

We use standard graph notation, including $u \to v$ to denote an arc from $u$ to $v$. We say that a graph is \emph{simple} and \emph{2-out-regular} when every vertex has exactly two outgoing arcs (with no bounds on incoming arcs) to two distinct vertices other than itself (i.e.~there are no loops and no parallel arcs). A \emph{Hamiltonian cycle} is a directed cycle that visits every vertex exactly once.

\providecommand\hamcycle{\textup{\textsc{2-Out Ham-Cycle}}}
% one before => after pair: #1 = board before, #2 = arrow (tikz path), #3 = board after.
% The pair is drawn at full size and shrunk only if it is wider than the column.
\ifcsname tlpairbox\endcsname\else\newsavebox\tlpairbox\fi
\providecommand\tlpair[3]{%
  \sbox\tlpairbox{%
    \begin{tikzpicture}[baseline=(current bounding box.center)]
      \tlgrid[cell=6mm]{#1}
      \draw[tl/move] #2;
    \end{tikzpicture}%
    \;$\Rightarrow$\;%
    \tlboard[cell=6mm]{#3}}%
  \ifdim\wd\tlpairbox>\linewidth
    \resizebox{\linewidth}{!}{\usebox\tlpairbox}%
  \else
    \usebox\tlpairbox
  \fi}

% panel labels (a), (b), ... that \ref can point to
\ifcsname c@tlpanel\endcsname\else\newcounter{tlpanel}\fi
\renewcommand\thetlpanel{(\alph{tlpanel})}
\providecommand\tlpanellabel{\par\smallskip\refstepcounter{tlpanel}\thetlpanel}

\begin{figure}[!ht]
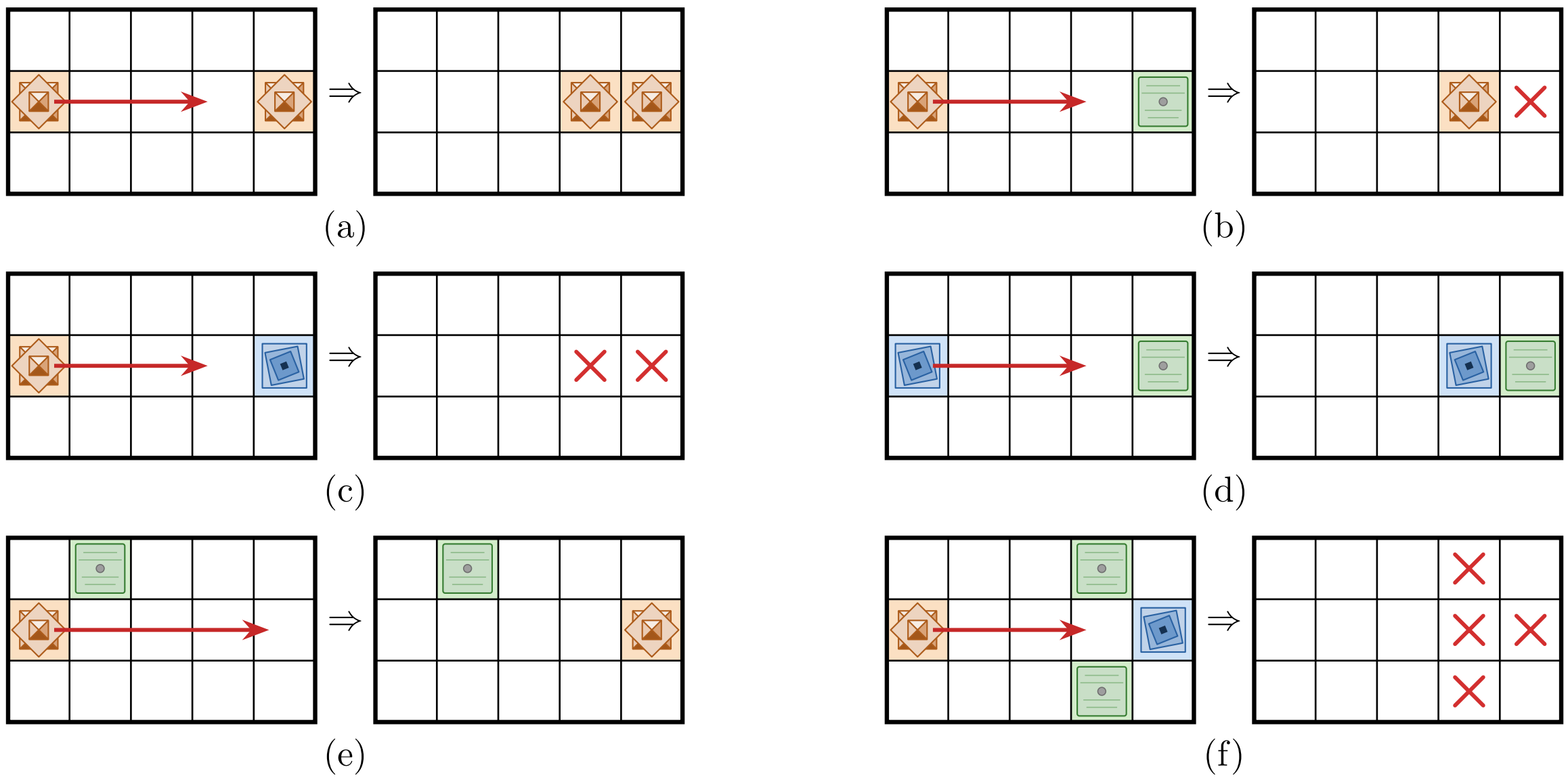

  \centering\setcounter{tlpanel}{0}%
  \begin{minipage}[b]{0.48\textwidth}\centering
    \tlpair{....., P...P, .....}{(tl-1-0) -- (tl-1-3)}{....., ...PP, .....}
    \tlpanellabel\label{fig:rules-slide}
  \end{minipage}\hfill
  \begin{minipage}[b]{0.48\textwidth}\centering
    \tlpair{....., P...F, .....}{(tl-1-0) -- (tl-1-3)}{....., ...Px, .....}
    \tlpanellabel\label{fig:rules-flat}
  \end{minipage}

  \medskip
  \begin{minipage}[b]{0.48\textwidth}\centering
    \tlpair{....., P...K, .....}{(tl-1-0) -- (tl-1-3)}{....., ...xx, .....}
    \tlpanellabel\label{fig:rules-stack}
  \end{minipage}\hfill
  \begin{minipage}[b]{0.48\textwidth}\centering
    \tlpair{....., K...F, .....}{(tl-1-0) -- (tl-1-3)}{....., ...KF, .....}
    \tlpanellabel\label{fig:rules-stackflat}
  \end{minipage}

  \medskip
  \begin{minipage}[b]{0.48\textwidth}\centering
    \tlpair{.F..., P...., .....}{(tl-1-0) -- (tl-1-4)}{.F..., ....P, .....}
    \tlpanellabel\label{fig:rules-pass}
  \end{minipage}\hfill
  \begin{minipage}[b]{0.48\textwidth}\centering
    \tlpair{...F., P...K, ...F.}{(tl-1-0) -- (tl-1-3)}{...x., ...xx, ...x.}
    \tlpanellabel\label{fig:rules-many}
  \end{minipage}
  \caption{The rules of Tenebrae Lux. Each panel shows a board before and after one move;
    the arrow is the move and a red cross marks a destroyed piece.}
  \label{fig:rules}
\end{figure}

\noindent An \emph{$h\times w$ board} (or \emph{configuration}) is a grid of cells $(r,c)$ with $0\le r < h$ and $0\le c < w$, where rows are counted from the bottom and columns from the left (note that they are zero-indexed). Each cell is either empty or contains one of three kinds of pieces: a \emph{pyramid} \tlpiece{P}, a \emph{stack} \tlpiece{K}, or a \emph{flat} \tlpiece{F}. Pyramids and stacks are called \emph{movers}; flats never move. Two cells are \emph{neighbours} if they share a side.

In Tenebrae Lux, a \emph{move} consists of selecting a mover and one of the four cardinal directions (up, down, right, left). The mover slides through consecutive empty cells until the next cell is occupied or lies outside the board. It must travel at least one cell; a move is therefore illegal if the immediately adjacent cell in the chosen direction is occupied or outside the board. Pieces interact only at the mover's destination. If the mover is a pyramid, every neighbouring flat and stack is destroyed, and the pyramid is also destroyed if at least one neighbouring stack is destroyed. If the mover is a stack, every neighbouring pyramid is destroyed, and the stack is also destroyed if at least one neighbouring pyramid is destroyed. All removals are simultaneous. No other interactions happen: two pyramids, two stacks, or a stack and a flat may remain adjacent without affecting one another. Observe then that any destruction must involve at least one pyramid. We say that a board that admits a sequence of moves that leads to an empty board has a \emph{clearing sequence}.

Figure~\ref{fig:rules} illustrates these rules. Pieces of the same type do not interact (Figure~\ref{fig:rules}\ref{fig:rules-slide}). A moved pyramid destroys neighbouring flats \ref{fig:rules-flat}, while a pyramid and a stack destroy one another when either moves into a neighbouring cell of the other \ref{fig:rules-stack}. A stack does not destroy neighbouring flats \ref{fig:rules-stackflat}. Only the final position matters: a mover that merely slides past another piece leaves it unaffected \ref{fig:rules-pass}. A move can also destroy several pieces simultaneously. In \ref{fig:rules-many}, the moved pyramid destroys both neighbouring flats and the neighbouring stack, and is itself destroyed because of the stack. Figure~\ref{fig:ingame-solution} illustrates a complete clearing sequence.

\begin{prop}\label{prop:membership}
\textup{\textsc{Tenebrae Lux}} is in PSPACE. For every fixed constant $k$, its restriction to boards with at most $k$ movers is in NP, and its restriction to boards with at most $k$ pieces in total is in P.
\end{prop}

\begin{proof}
Let $N=hw$, and assume that the board is encoded explicitly. There are at most $4^N$ configurations, since each cell is either empty or contains one of the three types of pieces. If a clearing sequence exists, then one exists that never repeats a configuration, and hence uses fewer than $4^N$ moves. A nondeterministic algorithm can guess such a sequence while storing only the current board and a counter bounded by $4^N$. Both require $O(N)$ space, and each move can be simulated in polynomial time. Thus \textsc{Tenebrae Lux} belongs to NPSPACE, which equals PSPACE by Savitch's theorem.

Now suppose that the initial board contains $t$ pieces, at most $k$ of which are movers, and fix a shortest clearing sequence. Call a move \emph{destroying} if it removes at least one piece. Since pieces are never introduced, there are at most $t$ destroying moves. These divide the remaining moves into at most $t+1$ phases containing no destructions. Within each phase, the flats remain unchanged, as do the numbers and types of surviving movers. The configuration is therefore determined by the positions of at most $k$ movers, giving at most $N^k$ configurations. A shortest clearing sequence cannot repeat a configuration, so each phase contains at most $N^k-1$ moves. Consequently, the entire sequence has length at most $(t+1)(N^k-1)+t < (t+1)N^k$.
Since $t\le N$ and $k$ is fixed, this bound is polynomial in the input size. Encoding each move by its source cell and direction gives a polynomial-size certificate that can be verified by simulating the sequence. Hence the restricted problem belongs to NP.

Finally, suppose that the board contains at most $k$ pieces in total. Each piece either occupies one of the $N$ cells or has been destroyed, so there are at most $(N+1)^k$ configurations. Each configuration has at most $4k$ outgoing moves, and each move can be simulated in polynomial time. We can therefore determine whether the empty board is reachable by exploring the configuration graph. For fixed $k$, this takes polynomial time in the size of the explicitly encoded board.
\end{proof}

% ---------------------------------------------------------------------------
% A solution of the in-game board.  Needs: tenebraelux.sty (v1.5).
% Each panel shows the board AFTER the numbered move.  The panels are sized to
% fill the text width (four per row), up to a maximum cell size of 8mm.
% ---------------------------------------------------------------------------
\ifcsname tlmovecell\endcsname\else\newlength\tlmovecell\fi
% #1 = label under the board, #2 = board, #3 = arrow path (may be empty)
\providecommand\tlmove[3]{%
  \begin{tikzpicture}[baseline=(current bounding box.north)]
    \tlgrid[cell=\tlmovecell]{#2}
    \if\relax\detokenize{#3}\relax\else
      \draw[tl/move, -{Stealth[length=\tlmoveback]}, shorten >=\tlmoveshorten] #3;
    \fi
    \node[below=0.6mm, font=\small] at (current bounding box.south) {#1};
  \end{tikzpicture}}

\begin{figure}[!ht]
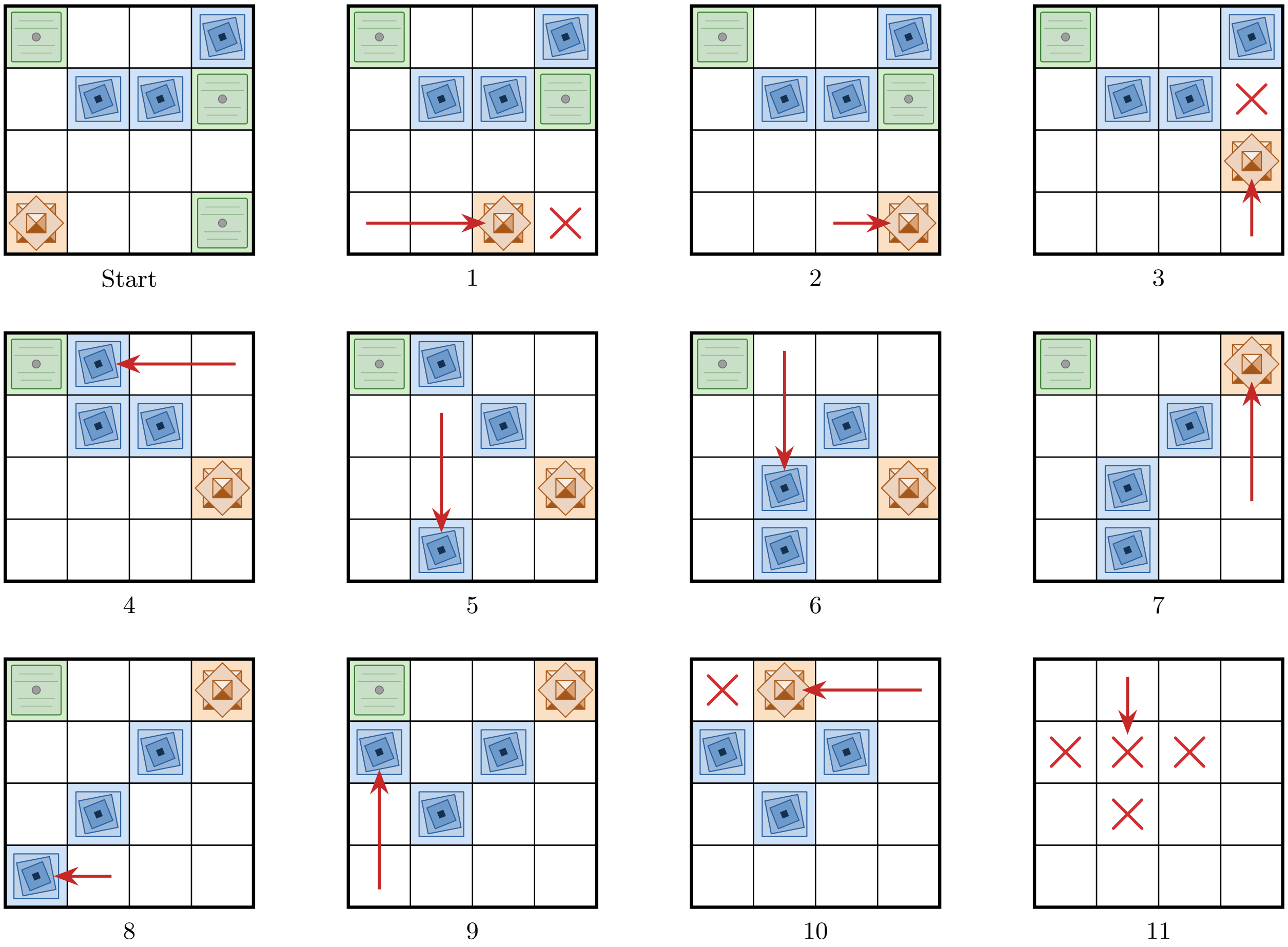

  \centering
  % four boards of four cells per row, with room for the gaps between them
  \setlength\tlmovecell{\dimexpr(\linewidth-4.5em)/16\relax}%
  \ifdim\tlmovecell>8mm \setlength\tlmovecell{8mm}\fi
  % arrows start 0.4 of a cell behind the centre of the starting cell
  \edef\tlmoveback{\the\dimexpr 4\tlmovecell/10\relax}%
  \edef\tlmoveshorten{\the\dimexpr 28\tlmovecell/100\relax}%
    \tlmove{Start}{F..K, .KKF, ...., P..F}{}\hfill
    \tlmove{1}{F..K, .KKF, ...., ..Px}{([shift={(-\tlmoveback,0)}]tl-3-0) -- (tl-3-2)}\hfill
    \tlmove{2}{F..K, .KKF, ...., ...P}{([shift={(-\tlmoveback,0)}]tl-3-2) -- (tl-3-3)}\hfill
    \tlmove{3}{F..K, .KKx, ...P, ....}{([shift={(0,-\tlmoveback)}]tl-3-3) -- (tl-2-3)}\par\bigskip
    \tlmove{4}{FK.., .KK., ...P, ....}{([shift={(\tlmoveback,0)}]tl-0-3) -- (tl-0-1)}\hfill
    \tlmove{5}{FK.., ..K., ...P, .K..}{([shift={(0,\tlmoveback)}]tl-1-1) -- (tl-3-1)}\hfill
    \tlmove{6}{F..., ..K., .K.P, .K..}{([shift={(0,\tlmoveback)}]tl-0-1) -- (tl-2-1)}\hfill
    \tlmove{7}{F..P, ..K., .K.., .K..}{([shift={(0,-\tlmoveback)}]tl-2-3) -- (tl-0-3)}\par\bigskip
    \tlmove{8}{F..P, ..K., .K.., K...}{([shift={(\tlmoveback,0)}]tl-3-1) -- (tl-3-0)}\hfill
    \tlmove{9}{F..P, K.K., .K.., ....}{([shift={(0,-\tlmoveback)}]tl-3-0) -- (tl-1-0)}\hfill
    \tlmove{10}{xP.., K.K., .K.., ....}{([shift={(\tlmoveback,0)}]tl-0-3) -- (tl-0-1)}\hfill
    \tlmove{11}{...., xxx., .x.., ....}{([shift={(0,\tlmoveback)}]tl-0-1) -- (tl-1-1)}
  \caption{A shortest clearing sequence of a solvable board, consisting of eleven moves. Each panel shows the board after the numbered move.}
  \label{fig:ingame-solution}
\end{figure}

\section{Proof of Hardness}\label{sec:hardness}

This section is devoted to the proof of Theorem~\ref{thm:main}. Again, we
reduce from the following restricted version of \textsc{Hamiltonian Cycle},
which is NP-complete by Rusu~\cite[Corollary~3]{rusu2022hamiltonian}. The relevant digraphs in \cite{rusu2022hamiltonian} have no multiple arcs by definition, and those constructed in the proof of
Corollary~3 have no loops.

\begin{center}
\problemStatement{\textsc{2-Out-Hamiltonian Cycle} (2OHC)}
  {A simple 2-out-regular graph $G$.}
  {Does there exist a Hamiltonian cycle in $G$?}
\end{center}

\noindent We first describe how to transform an instance $G$ of 2OHC into a board $B$ of
\textsc{Tenebrae Lux} (reduction), and we prove some properties of $B$ that
both directions of the proof use. We then show that if $G$ has a Hamiltonian
cycle, then $B$ has a clearing sequence (completeness), and finally that if $B$
has a clearing sequence, then $G$ has a Hamiltonian cycle (soundness).

\subsection{Reduction of a 2OHC Instance}\label{subsec:reduction}

Consider a simple 2-out-regular digraph $G=(V,E)$. Let $|V|=n$, and choose a
board with $h=8n+8$ rows and $w=4n^2+4n+8$ columns. We index the cells as (row
number, column number) as in Subsection \ref{ss:prelim}.

We put a stack at $(0,0)$ and flats at $(0,1)$ and $(1,0)$, which we call the
\emph{blocking flats}. While both blocking flats are present, the stack cannot move.

Define an ordering on the vertices such that $V=\{v_1,\ldots,v_n\}$, and write
$C_i=3+4i(n+1)$ for $i\in[1,n]$; these columns increase with $i$, and
consecutive ones are $4(n+1)$ apart. For each $v_i$, let $v_i\to v_j$ and
$v_i\to v_k$ be its two outgoing arcs, with $j<k$. Throughout this section, $j$
and $k$ denote these two indices for whichever vertex $v_i$ is being considered.
Let
\[
  A_i=4n+3+4i,\qquad B_i=4i+1,\qquad D_i=C_j+4i.
\]
We call row $A_i$ the \emph{arrival row} of $v_i$, row $B_i$ its
\emph{departure row}, column $C_i$ its \emph{arrival column} and column $D_i$ its
\emph{departure column}. Since $0<4i<4(n+1)$ and $j<k$, we have
\begin{equation}\label{eq:D}
  C_j<D_i<C_{j+1}\le C_k .
\end{equation}
Since $G$ has no loops, $i\neq j$, and comparing $C_i$ with \eqref{eq:D} gives
\begin{equation}\label{eq:side}
  C_i<D_i \text{ if } i<j, \qquad D_i<C_i \text{ if } i>j .
\end{equation}

\noindent For each $v_i$, we place the five flats in Table~\ref{tab:flats}: two flats
$v_i^-$ and $v_i^+$, an intermediate flat $y(v_i)$ that lets the pyramid go
from $v_i^-$ to $v_i^+$, and a flat for each outgoing arc of $v_i$. Each flat
lies one cell beyond a cell where it is meant to stop the pyramid. By
\eqref{eq:side}, $y(v_i)$ lies on the side of column $D_i$ away from column
$C_i$, and by \eqref{eq:D}, the two $z$ flats lie on the sides of columns $C_j$
and $C_k$ away from column $D_i$ (see Figure~\ref{fig:edge-closeup}). All flats other
than the blocking flats lie in rows $B_1-1=4$ to $A_n+1=h-4$ and in columns
$C_1-1=4n+6$ to $C_n+1=w-4$.

% ---------------------------------------------------------------------------
% Close-up of the flats for the arcs of one vertex v_i (Section 2.1, with (0,0) in
% the bottom left).  Needs: tenebraelux.sty (v1.5).
% The view keeps 9 rows and 9 columns of the board; the grey dotted strips
% (one gap row and one gap column between groups) stand for many omitted lines,
% which can contain flats of other vertices.  Every drawn cell is exact.
% Shown rows, top to bottom: A_k+2, A_k+1, A_k | A_j+2, A_j+1, A_j | B_i+1, B_i, B_i-1.
% Shown columns, left to right: C_j-1, C_j, C_j+1 | D_i-1, D_i, D_i+1 | C_k-1, C_k, C_k+1,
% where A_i = 4n+3+4i, B_i = 4i+1, C_x = 3+4x(n+1) and D_i = C_j+4i (Section 2.1).
% Checked against the simulator on the instance of Figure~\ref{fig:reduction-example}
% (i = 2, j = 1, k = 3).
% ---------------------------------------------------------------------------
\ifcsname tledgecell\endcsname\else\newlength\tledgecell\fi
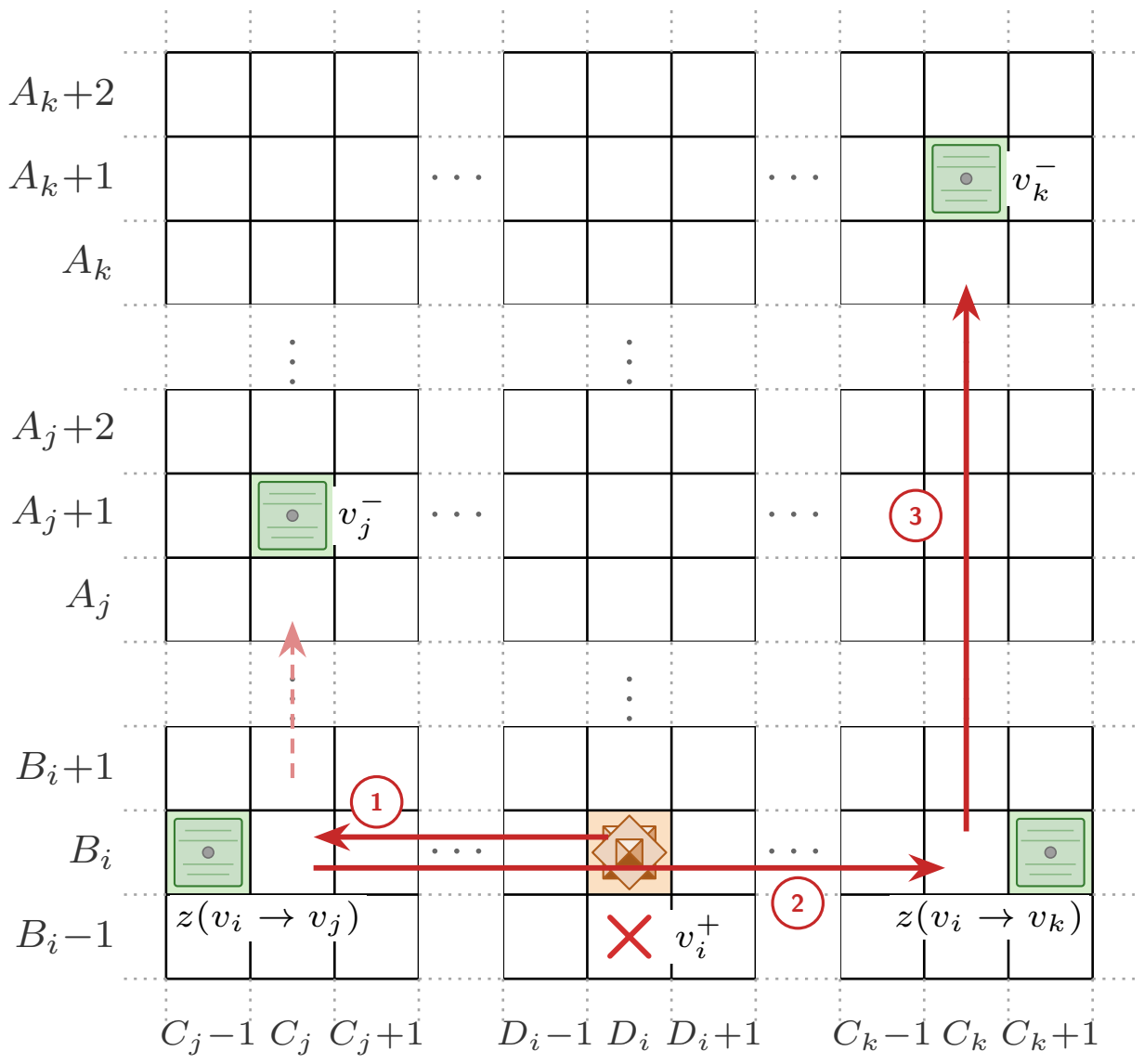
\begin{figure}[!ht]
  \centering
  \setlength\tledgecell{6mm}%
  \begin{tikzpicture}[
      gapline/.style={draw=black!35, line width=0.5pt, dash pattern=on 0.6pt off 1.4pt},
      flatlabel/.style={font=\scriptsize, inner sep=1pt, fill=white},
      coordlabel/.style={font=\footnotesize, inner sep=1pt, text=black!80}]
    % the selected rows and columns (rows 3 and 7, columns 3 and 7 are gaps)
    \tlgrid[cell=\tledgecell, frame=false]{
      ...........,
      .........F.,
      ...........,
      ...........,
      ...........,
      .F.........,
      ...........,
      ...........,
      ...........,
      F....P....F,
      .....x.....}
    \begin{scope}[x=\tledgecell, y=-\tledgecell]
      % blank out the gap strips ...
      \fill[white] (3,-0.6) rectangle (4,11.6);
      \fill[white] (7,-0.6) rectangle (8,11.6);
      \fill[white] (-0.6,3) rectangle (11.6,4);
      \fill[white] (-0.6,7) rectangle (11.6,8);
      % ... continue the grid lines through them as dotted lines ...
      \foreach \y in {0,1,2,3,4,5,6,7,8,9,10,11} {
        \draw[gapline] (3,\y) -- (4,\y);
        \draw[gapline] (7,\y) -- (8,\y);
      }
      \foreach \x in {0,1,2,3,4,5,6,7,8,9,10,11} {
        \draw[gapline] (\x,3) -- (\x,4);
        \draw[gapline] (\x,7) -- (\x,8);
      }
      % ... and past the edges of the view, which is only part of the board
      \foreach \y in {0,1,2,3,4,5,6,7,8,9,10,11} {
        \draw[gapline] (-0.5,\y) -- (0,\y);
        \draw[gapline] (11,\y) -- (11.5,\y);
      }
      \foreach \x in {0,1,2,3,4,5,6,7,8,9,10,11} {
        \draw[gapline] (\x,-0.5) -- (\x,0);
        \draw[gapline] (\x,11) -- (\x,11.5);
      }
      % ellipses in the gaps
      \foreach \y in {1.5,5.5,9.5} {
        \node[text=black!60] at (3.5,\y) {$\cdots$};
        \node[text=black!60] at (7.5,\y) {$\cdots$};
      }
      \foreach \x in {1.5,5.5,9.5} {
        \node[text=black!60] at (\x,3.5) {$\vdots$};
        \node[text=black!60] at (\x,7.5) {$\vdots$};
      }
    \end{scope}
    % moves for l = k: left, right, up
    \draw[tl/move] ([yshift=1.1mm]tl-9-5) -- ([yshift=1.1mm]tl-9-1);
    \draw[tl/move] ([yshift=-1.1mm]tl-9-1) -- ([yshift=-1.1mm]tl-9-9);
    \draw[tl/move] (tl-9-9) -- (tl-2-9);
    \node[tl/step] at ([yshift=3.6mm]tl-9-2) {1};
    \node[tl/step] at ([yshift=-3.6mm]tl-9-7) {2};
    \node[tl/step] at ([xshift=-3.6mm]tl-5-9) {3};
    % the alternative: had z(v_i -> v_k) been destroyed first, the pyramid would
    % have ended in column C_j and continued up to v_j^-
    \draw[tl/move, densely dashed, draw=tlMoveLine!55, line width=0.8pt] ([yshift=3.8mm]tl-9-1) -- (tl-6-1);
    % flat labels
    \node[flatlabel, anchor=north west] at ([xshift=-2.8mm, yshift=-3mm]tl-9-0) {$z(v_i \to v_j)$};
    \node[flatlabel, anchor=north east] at ([xshift=2.8mm, yshift=-3mm]tl-9-10) {$z(v_i \to v_k)$};
    \node[flatlabel, anchor=west] at ([xshift=2.8mm]tl-5-1) {$v_j^-$};
    \node[flatlabel, anchor=west] at ([xshift=2.8mm]tl-1-9) {$v_k^-$};
    \node[flatlabel, anchor=west] at ([xshift=2.8mm]tl-10-5) {$v_i^+$};
    % row names (A_i = 4n+3+4i, B_i = 4i+1, as in Section 2.1)
    \foreach \r/\t in {0/{A_k{+}2}, 1/{A_k{+}1}, 2/{A_k},
                       4/{A_j{+}2}, 5/{A_j{+}1}, 6/{A_j},
                       8/{B_i{+}1}, 9/{B_i}, 10/{B_i{-}1}}
      \node[coordlabel, anchor=east] at ([xshift=-6.2mm]tl-\r-0) {$\t$};
    % column names (C_x = 3+4x(n+1), D_i = C_j+4i, as in Section 2.1)
    \foreach \c/\t in {0/{C_j{-}1}, 1/{C_j}, 2/{C_j{+}1},
                       4/{D_i{-}1}, 5/{D_i}, 6/{D_i{+}1},
                       8/{C_k{-}1}, 9/{C_k}, 10/{C_k{+}1}}
      \node[coordlabel, font=\scriptsize, anchor=north] at ([yshift=-5.6mm]tl-10-\c) {$\t$};
  \end{tikzpicture}
  \caption{Placement of flats for the arcs $v_i \to v_j$ and $v_i \to v_k$ with $j < k$. Here, traversing $G$ from $v_i$ to $v_k$, for instance, corresponds to the pyramid moving towards $z(v_i \to v_j)$ (on the left), destroying it, then moving towards $z(v_i \to v_k)$ (on the right), destroying it, and finally moving up to $v^-_k$, which corresponds to an incoming arc of $v_k$ in $G$.}
  \label{fig:edge-closeup}
\end{figure}

\begin{table}[ht]
\centering
\begin{tabular}{@{}lll@{}}
\hline
Flat & Cell & Stops the pyramid at\\
\hline
$v_i^-$ & $(A_i+1,\ C_i)$ & $(A_i,\ C_i)$, moving up column $C_i$\\
$y(v_i)$, if $i<j$ & $(A_i,\ D_i+1)$ & $(A_i,\ D_i)$, moving right along row $A_i$\\
$y(v_i)$, if $i>j$ & $(A_i,\ D_i-1)$ & $(A_i,\ D_i)$, moving left along row $A_i$\\
$v_i^+$ & $(B_i-1,\ D_i)$ & $(B_i,\ D_i)$, moving down column $D_i$\\
$z(v_i\to v_j)$ & $(B_i,\ C_j-1)$ & $(B_i,\ C_j)$, moving left along row $B_i$\\
$z(v_i\to v_k)$ & $(B_i,\ C_k+1)$ & $(B_i,\ C_k)$, moving right along row $B_i$\\
\hline
\end{tabular}
\caption{The flats placed for a vertex $v_i$, where $v_i\to v_j$ and
$v_i\to v_k$ are its outgoing arcs with $j<k$.}
\label{tab:flats}
\end{table}

\noindent Finally, choose an arbitrary $\rho\in[1,n]$. Remove $v_\rho^+$ and $y(v_\rho)$
from the board, and place the pyramid at $(B_\rho,\ D_\rho)$, the cell
immediately above where $v_\rho^+$ was. This completes the construction of $B$
(see Figure~\ref{fig:reduction-example} for an example).

\noindent We use the following terms for moves. A move slides the moved piece until the
next cell in its direction is occupied or lies outside the board. In the first
case, we say that the moved piece \emph{hits} the piece in that cell; it then
stops next to it. A moved pyramid destroys every flat next to its destination.
In particular, it destroys every flat that it hits, but it may also destroy
flats that it does not hit.

\subsection{Properties of the Board}\label{subsec:properties}

We now collect some properties of $B$ that both directions of the proof use.
Let $\mathcal{N}$ be the set of flats $v_i^-$ and $v_i^+$. A \emph{line} is a
row or a column. We call rows $0,1,2,h-1$ and columns $0,1,2,w-1$
\emph{outer}, and we call a cell an \emph{outer crossing} if both its row and its
column are outer. Recall that the stack never destroys flats. It cannot move
while both blocking flats are present, and otherwise it slides in the same way
as the pyramid. If the pyramid and the stack become neighbours, whichever of
them moved, both are destroyed; no movers then remain, so the board has been
cleared if and only if it is now empty.

\begin{lemma}\label{lem:contents}
The following four statements hold.
\begin{enumerate}
  \item Every flat of $\mathcal{N}$ lies in a row $\equiv 0 \pmod 4$ and a
        column $\equiv 3 \pmod 4$. Every $y$ and $z$ flat lies in a column
        $\equiv 0$ or $2 \pmod 4$, next to exactly one column $\equiv 3
        \pmod 4$: column $C_\ell$ for $z(v_i\to v_\ell)$, and column $D_i$ for
        $y(v_i)$.
  \item Row $B_i$ contains only $z(v_i\to v_j)$ and $z(v_i\to v_k)$, and row
        $A_i$ contains only $y(v_i)$. Row $B_i-1$ contains only $v_i^+$, and row
        $A_i+1$ contains only $v_i^-$. Rows $B_\rho-1$ and $A_\rho$ are empty, and
        so are rows $B_i+1$, $A_i-1$ and $A_i+2$.
  \item Column $C_i$ contains only $v_i^-$, and column $D_i$ contains only
        $v_i^+$; column $D_\rho$ is empty. The columns $C_1,\ldots,C_n$ and
        $D_1,\ldots,D_n$ are all distinct.
  \item The outer rows and columns contain no flats other than the blocking
        flats.
\end{enumerate}
\end{lemma}

\begin{proof}
Statements (1), (2) and (4) follow directly from Table~\ref{tab:flats}. For (3),
statement (1) shows that no $y$ or $z$ flat lies in a column $\equiv 3 \pmod 4$,
and the blocking flats lie in columns $0$ and $1$, so it suffices to show that
the columns $C_i$ and $D_i$ are distinct. We have $(C_i-3)/4=i(n+1)\equiv
0\pmod{n+1}$ and $(D_i-3)/4=j(n+1)+i\equiv i\pmod{n+1}$, where $1\le i\le n$. So
no $C$ column equals a $D$ column, two $D$ columns with different $i$ differ,
and the $C$ columns are distinct as they increase with $i$.
\end{proof}

By Lemma~\ref{lem:contents}(2), no two flats share a cell. No flat is next to
the pyramid: the cell $(B_\rho-1,\ D_\rho)$ below it was emptied by removing
$v_\rho^+$, row $B_\rho+1$ is empty, and since $C_j<D_\rho<C_k$ are all
$\equiv 3\pmod 4$, the $z$ flats of $v_\rho$, which are the only flats in row
$B_\rho$, lie at least five columns away. So $B$ is a valid instance of
\textsc{Tenebrae Lux}. It has $5n$ flats ($2n-1$ in $\mathcal{N}$, $n-1$ $y$
flats, $2n$ $z$ flats and the two blocking flats), one pyramid and one stack, on
$(8n+8)(4n^2+4n+8)=\Theta(n^3)$ cells, and it can clearly be built in polynomial
time.

\begin{lemma}\label{lem:stack}
The stack is always at an outer crossing. In particular, every cell next to the
stack lies on an outer line and is not next to any flat other than the blocking
flats.
\end{lemma}

\begin{proof}
By Lemma~\ref{lem:contents}(4), the only flats on outer lines are the blocking
flats. So a move of the stack along an outer line ends at the end of that line,
next to a blocking flat that it hits, or next to the pyramid, in which case the
stack is destroyed. Each end of an outer line lies on an outer line at the edge
of the board, and every cell next to a blocking flat lies in rows $0$ to $2$ and
columns $0$ to $2$, so in both cases the stack ends at an outer crossing. The
stack starts at the outer crossing $(0,0)$, and every line through an outer
crossing is outer, so by induction the stack is always at an outer crossing.

Every cell next to an outer crossing shares its row or its column, so it lies on
an outer line. The flats other than the blocking flats lie in rows $4$ to $h-4$
and columns $4n+6$ to $w-4$, so none of them is next to a cell on an outer
line.
\end{proof}

\begin{lemma}\label{lem:onekill}
Whenever a move of the pyramid ends, it destroys at most one flat. Moreover, if
it destroys a flat other than a blocking flat, then it hits that flat.
\end{lemma}

\begin{proof}
Let the move end at $s$. Suppose first that the pyramid destroys a flat $f$
other than a blocking flat, so that $f$ is next to $s$. Then $s$ is not on the
edge of the board, since $f$ lies in rows $4$ to $h-4$ and columns $4n+6$ to
$w-4$, and $s$ is not next to the stack, by Lemma~\ref{lem:stack}. So the
pyramid hit a flat $g$ next to $s$. The cell on the other side of $s$ from $g$
is the one the pyramid came from, so it is empty. Hence every other flat next to
$s$ lies in a row next to the row of $g$ and in a column next to the column of
$g$.

No two flats lie in adjacent rows and adjacent columns, apart from the two
blocking flats. Indeed, the only other pairs of adjacent rows that both contain
flats are rows $B_i-1$ and $B_i$, which contain $v_i^+$ and the $z$ flats of
$v_i$, and rows $A_i$ and $A_i+1$, which contain $y(v_i)$ and $v_i^-$. By
Lemma~\ref{lem:contents}(1), the only column $\equiv 3\pmod 4$ next to
$z(v_i\to v_\ell)$ is $C_\ell$, which differs from $D_i$, and the only one next
to $y(v_i)$ is $D_i$, which differs from $C_i$. So neither $z$ flat of $v_i$ is
in a column next to that of $v_i^+$, and $y(v_i)$ is not in a column next to
that of $v_i^-$. As $f$ is not a blocking flat, it follows that $f=g$, and that
the pyramid destroys no other flat.

Now suppose that the pyramid destroys both blocking flats, so that both are
present before the move. The stack cannot move while both are present, so it is
still at $(0,0)$. The only cells next to both blocking flats are $(0,0)$ and
$(1,1)$, so the pyramid ends at $(1,1)$, which is not next to the stack, and it
survives the move. However, in a move that it survives, the pyramid can only stop
in row $1$ by moving along row $1$ or by hitting a flat in row $0$ or $2$ of the
same column, and the only such flat, $(0,1)$, is in column $1$. Similarly, it can
only stop in column $1$ by moving along column $1$ or by hitting a flat in column
$0$ or $2$ of the same row, and the only such flat, $(1,0)$, is in row $1$. So
the pyramid can only stop in row $1$ or column $1$ if it is already in row $1$ or
column $1$. As it starts in neither, it never stops at $(1,1)$, a contradiction. 
\end{proof}

% ---------------------------------------------------------------------------
% The reduction on the smallest instance: the complete digraph on three vertices.
% Needs: tenebraelux.sty (v1.5).  Coordinates are exactly those of Section 2.1
% (n = 3, r = 32 rows, c = 56 columns, rho = 1); the matrix below has the top row
% (row r-1 of the paper) first.  Generated and checked by simulation.
% ---------------------------------------------------------------------------
\ifcsname tlredcell\endcsname\else\newlength\tlredcell\fi
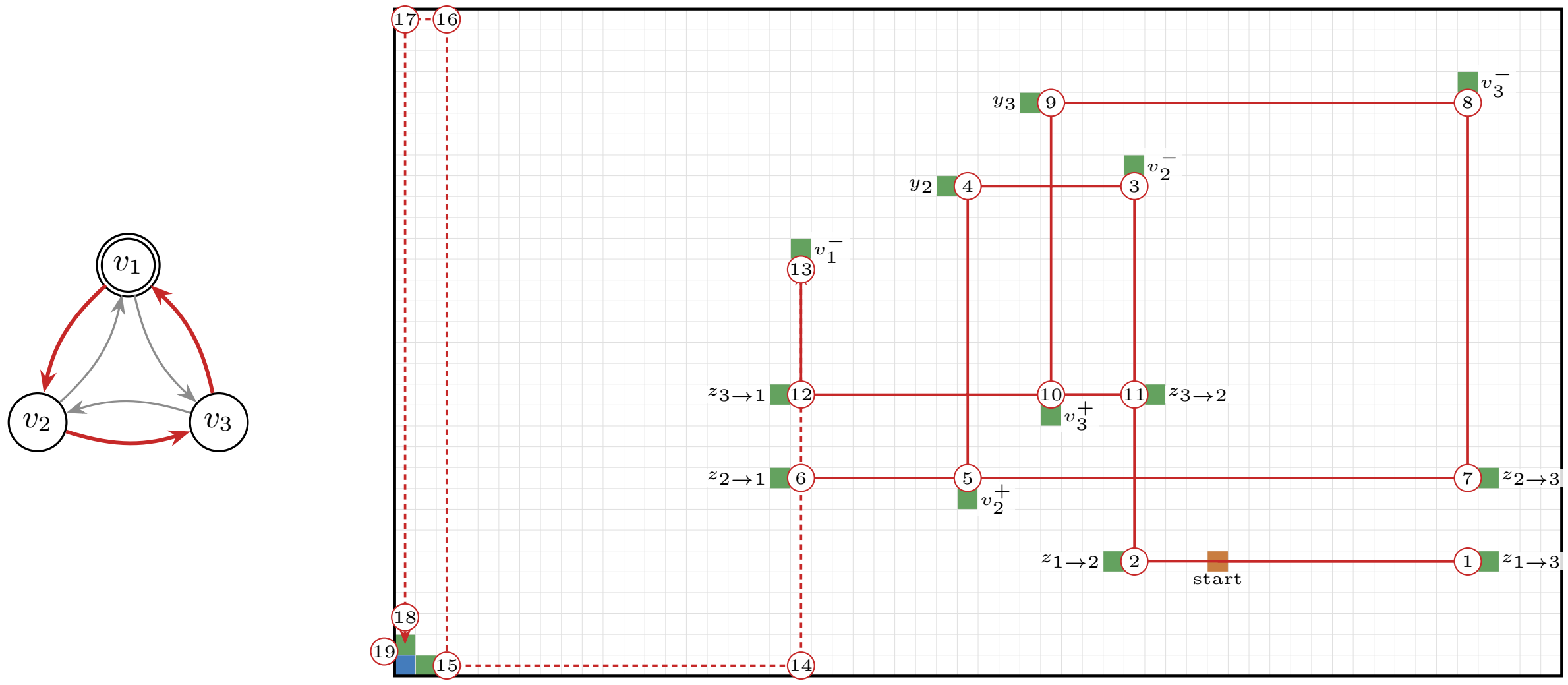
\begin{figure}[!ht]
  \centering
  % the graph
  \begin{minipage}[c]{0.22\linewidth}\centering
    \begin{tikzpicture}[
        vtx/.style={circle, draw, line width=0.6pt, minimum size=5.8mm, inner sep=0pt, font=\small},
        arc/.style={-{Stealth[length=2mm]}, line width=0.6pt, black!45},
        ham/.style={-{Stealth[length=2.2mm]}, line width=1.1pt, tlMoveLine}]
      \node[vtx, double, double distance=0.8pt] (v1) at (90:1.05)  {$v_1$};
      \node[vtx] (v2) at (210:1.05) {$v_2$};
      \node[vtx] (v3) at (330:1.05) {$v_3$};
      \draw[ham] (v1) to[bend right=18] (v2);
      \draw[ham] (v2) to[bend right=18] (v3);
      \draw[ham] (v3) to[bend right=18] (v1);
      \draw[arc] (v2) to[bend right=18] (v1);
      \draw[arc] (v3) to[bend right=18] (v2);
      \draw[arc] (v1) to[bend right=18] (v3);
    \end{tikzpicture}
  \end{minipage}\hfill
  % the board
  \begin{minipage}[c]{0.76\linewidth}\centering
    \setlength\tlredcell{\dimexpr\linewidth/60\relax}%
    \edef\tlredoff{\the\dimexpr 45\tlredcell/100\relax}%
    \edef\tlrednudge{\the\dimexpr 32\tlredcell/100\relax}%
    \edef\tlredcellw{\the\dimexpr \tlredcell\relax}%
    % stronger fills, so that single cells stay visible at this size
    \colorlet{tlPyramidFill}{tlPyramidLine!85}\colorlet{tlStackFill}{tlStackLine!90}\colorlet{tlFlatFill}{tlFlatLine!80}%
    \begin{tikzpicture}[
        tl/grid/.style={draw=black!12, line width=0.15pt},
        tl/frame/.style={draw=black, line width=0.8pt},
        tl/redstep/.style={circle, draw=tlMoveLine, fill=white, line width=0.4pt, inner sep=0pt,
          minimum size=2.7mm, font=\fontsize{4.6}{5}\selectfont},
        tl/redlabel/.style={font=\tiny, inner sep=0.6pt, fill=white, fill opacity=0.85, text opacity=1}]
      \tlgrid[cell=\tlredcell, glyphs=false]{
      ........................................................,
      ........................................................,
      ........................................................,
      ...................................................F....,
      ..............................F.........................,
      ........................................................,
      ........................................................,
      ...................................F....................,
      ..........................F.............................,
      ........................................................,
      ........................................................,
      ...................F....................................,
      ........................................................,
      ........................................................,
      ........................................................,
      ........................................................,
      ........................................................,
      ........................................................,
      ..................F.................F...................,
      ...............................F........................,
      ........................................................,
      ........................................................,
      ..................F.................................F...,
      ...........................F............................,
      ........................................................,
      ........................................................,
      ..................................F....P............F...,
      ........................................................,
      ........................................................,
      ........................................................,
      F.......................................................,
      KF......................................................}
      \node[tl/redlabel, anchor=west] at ([xshift=\tlredoff]tl-11-19) {$v_{1}^{-}$};
      \node[tl/redlabel, anchor=east] at ([xshift=-\tlredoff]tl-26-34) {$z_{1\to 2}$};
      \node[tl/redlabel, anchor=west] at ([xshift=\tlredoff]tl-26-52) {$z_{1\to 3}$};
      \node[tl/redlabel, anchor=west] at ([xshift=\tlredoff]tl-7-35) {$v_{2}^{-}$};
      \node[tl/redlabel, anchor=west] at ([xshift=\tlredoff]tl-23-27) {$v_{2}^{+}$};
      \node[tl/redlabel, anchor=east] at ([xshift=-\tlredoff]tl-8-26) {$y_{2}$};
      \node[tl/redlabel, anchor=east] at ([xshift=-\tlredoff]tl-22-18) {$z_{2\to 1}$};
      \node[tl/redlabel, anchor=west] at ([xshift=\tlredoff]tl-22-52) {$z_{2\to 3}$};
      \node[tl/redlabel, anchor=west] at ([xshift=\tlredoff]tl-3-51) {$v_{3}^{-}$};
      \node[tl/redlabel, anchor=west] at ([xshift=\tlredoff]tl-19-31) {$v_{3}^{+}$};
      \node[tl/redlabel, anchor=east] at ([xshift=-\tlredoff]tl-4-30) {$y_{3}$};
      \node[tl/redlabel, anchor=east] at ([xshift=-\tlredoff]tl-18-18) {$z_{3\to 1}$};
      \node[tl/redlabel, anchor=west] at ([xshift=\tlredoff]tl-18-36) {$z_{3\to 2}$};
      \draw[tlMoveLine, line width=0.7pt, line join=round, -{Stealth[length=1.6mm]}]
        (tl-26-39) -- (tl-26-51) -- (tl-26-35) -- (tl-8-35) -- (tl-8-27) -- (tl-22-27) -- (tl-22-19) -- (tl-22-51) -- (tl-4-51) -- (tl-4-31) -- (tl-18-31) -- (tl-18-35) -- (tl-18-19) -- (tl-12-19);
      \draw[tlMoveLine, line width=0.7pt, line join=round, dash pattern=on 1.6pt off 1.2pt, -{Stealth[length=1.6mm]}]
        (tl-12-19) -- (tl-31-19) -- (tl-31-2) -- (tl-0-2) -- (tl-0-0) -- (tl-29-0) -- (tl-30-0);
      % the pyramid's position after each move
      \node[tl/redstep] at (tl-26-51) {1};
      \node[tl/redstep] at (tl-26-35) {2};
      \node[tl/redstep] at (tl-8-35) {3};
      \node[tl/redstep] at (tl-8-27) {4};
      \node[tl/redstep] at (tl-22-27) {5};
      \node[tl/redstep] at (tl-22-19) {6};
      \node[tl/redstep] at (tl-22-51) {7};
      \node[tl/redstep] at (tl-4-51) {8};
      \node[tl/redstep] at (tl-4-31) {9};
      \node[tl/redstep] at (tl-18-31) {10};
      \node[tl/redstep] at (tl-18-35) {11};
      \node[tl/redstep] at (tl-18-19) {12};
      \node[tl/redstep] at (tl-12-19) {13};
      \node[tl/redstep] at (tl-31-19) {14};
      \node[tl/redstep] at (tl-31-2) {15};
      \node[tl/redstep] at (tl-0-2) {16};
      \node[tl/redstep] at (tl-0-0) {17};
      \node[tl/redstep] at ([yshift=\tlrednudge]tl-29-0) {18};
      \node[tl/redstep] at ([yshift=-\tlrednudge, xshift=-\tlredcellw]tl-30-0) {19};
      \node[tl/redlabel, anchor=north] at ([yshift=-\tlredoff]tl-26-39) {start};
    \end{tikzpicture}
  \end{minipage}
  \caption{The reduction applied to the complete digraph on three vertices, which is 2-out-regular, with $\rho = 1$. \textbf{Left:} the graph, with $v_\rho$ circled twice and the Hamiltonian cycle $v_1 v_2 v_3 v_1$ in red. \textbf{Right:} the resulting board with $h = 32$ rows and $w = 56$ columns; orange is the pyramid, blue the stack and green the flats, and $y_i$ and $z_{a\to b}$ stand for $y(v_i)$ and $z(v_a \to v_b)$. The flats $v_1^+$ and $y(v_1)$ are absent because $v_1 = v_\rho$. The red line is the clearing sequence from the proof of completeness, and each number marks where the pyramid stops after that move: the solid part follows the Hamiltonian cycle and destroys every labelled flat, and the dashed part destroys the blocking flats and finally hits the stack.}
  \label{fig:reduction-example}
\end{figure}

\subsection{Completeness (\texorpdfstring{$\Rightarrow$}{=>})}\label{subsec:completeness}

Assume that $G$ has a Hamiltonian cycle $v_{\sigma(1)}, v_{\sigma(2)}, \ldots, v_{\sigma(n)}, v_{\sigma(1)}$. We can rotate it to ensure that $\sigma(1) = \rho$. We will prove that $B$ has a clearing sequence. First, we require two lemmas to show how travelling along an outgoing arc of $G$ is represented by moves on the board, and similarly how arriving at and leaving a vertex is represented. By Lemma~\ref{lem:stack}, the stack is always at an outer crossing, so it is never in a row $A_i$ or $B_i$ or a column $C_i$ or $D_i$; in these lines, the only pieces other than the pyramid are therefore the flats given by Lemma~\ref{lem:contents}.

\begin{lemma}\label{lem:arc}
Let $\ell\in\{j,k\}$. Suppose that the pyramid is at $(B_i,\ D_i)$, the cell
immediately that of $v_i^+$, that $z(v_i\to v_j)$, $z(v_i\to v_k)$ and $v_\ell^-$
are on the board, and that $v_i^+$ is not. Then the pyramid can hit
$z(v_i\to v_j)$ and $z(v_i\to v_k)$, and then hit $v_\ell^-$ from the cell
immediately below it, $(A_\ell,\ C_\ell)$.
\end{lemma}

\begin{proof}
By Lemma~\ref{lem:contents}(2), the only flats in row $B_i$ are
$z(v_i\to v_j)$ at $(B_i,\ C_j-1)$ and $z(v_i\to v_k)$ at $(B_i,\ C_k+1)$, and by
\eqref{eq:D}, they lie to the left and to the right of the pyramid,
respectively. If $\ell=j$, move the pyramid right: it hits $z(v_i\to v_k)$ and
stops at $(B_i,\ C_k)$. Row $B_i$ now contains only $z(v_i\to v_j)$, so when the
pyramid moves left, it hits $z(v_i\to v_j)$ and stops at $(B_i,\ C_j)$. If
$\ell=k$, do the same in the other order: left to $(B_i,\ C_j)$, then right to
$(B_i,\ C_k)$.

In both cases, the pyramid is now at $(B_i,\ C_\ell)$. By
Lemma~\ref{lem:contents}(3), the only flat in column $C_\ell$ is $v_\ell^-$, at
$(A_\ell+1,\ C_\ell)$, which is above the pyramid. Moving up, the pyramid hits
$v_\ell^-$ and stops at $(A_\ell,\ C_\ell)$.
\end{proof}

\begin{lemma}\label{lem:vertex}
Let $i\neq\rho$. Suppose that the pyramid is at $(A_i,\ C_i)$, the cell
immediately below that of $v_i^-$, that $y(v_i)$ and $v_i^+$ are on the board, and that
$v_i^-$ is not. Then the pyramid can hit $y(v_i)$, and then hit $v_i^+$ from the
cell immediately above it, $(B_i,\ D_i)$.
\end{lemma}

\begin{proof}
By Lemma~\ref{lem:contents}(2), the only flat in row $A_i$ is $y(v_i)$. If
$i<j$, then $C_i<D_i$ by \eqref{eq:side}, and $y(v_i)$ is at $(A_i,\ D_i+1)$, to
the right of the pyramid; moving right, the pyramid hits $y(v_i)$ and stops at
$(A_i,\ D_i)$. If $i>j$, then $D_i<C_i$, and $y(v_i)$ is at $(A_i,\ D_i-1)$, to
the left of the pyramid; moving left, the pyramid hits $y(v_i)$ and stops at
$(A_i,\ D_i)$.

By Lemma~\ref{lem:contents}(3), the only flat in column $D_i$ is $v_i^+$, at
$(B_i-1,\ D_i)$, which is below the pyramid. Moving down, the pyramid hits
$v_i^+$ and stops at $(B_i,\ D_i)$.
\end{proof}

\noindent By applying Lemma~\ref{lem:arc} with $i=\sigma(1)=\rho$ and $\ell=\sigma(2)$,
the pyramid can hit both $z$ flats of $v_{\sigma(1)}$ and then
$v_{\sigma(2)}^-$. Then, by applying Lemma~\ref{lem:vertex} with $i=\sigma(2)$,
it can hit $y(v_{\sigma(2)})$ and $v_{\sigma(2)}^+$. We then alternately apply
Lemma~\ref{lem:arc} with $i=\sigma(t)$ and $\ell=\sigma(t+1)$, and
Lemma~\ref{lem:vertex} with $i=\sigma(t+1)$, for $t\in[2,n-1]$, and we finish
by applying Lemma~\ref{lem:arc} with $i=\sigma(n)$ and $\ell=\sigma(1)=\rho$.
By Lemma~\ref{lem:onekill}, each of these moves destroys only the flat that it
hits, and since every vertex appears exactly once in the cycle, the hypotheses
of each lemma hold when it is applied. This destroys all flats of $\mathcal{N}$
and all $y$ and $z$ flats, and the pyramid reaches $(A_\rho,\ C_\rho)$.

It now remains to clear the corner. We move the pyramid down, left, up, left, down and down, so that it stops at the following cells in turn:
\[
  (0,\ C_\rho),\quad (0,2),\quad (h-1,2),\quad (h-1,0),\quad (2,0),\quad (1,0).
\]
At $(0,2)$, the pyramid hits the blocking flat $(0,1)$ and destroys it,
and at $(2,0)$, it hits and destroys the blocking flat $(1,0)$. Finally, it hits
the stack and stops at $(1,0)$, so both are destroyed. The board is now empty,
so this sequence is clearing.

\subsection{Soundness (\texorpdfstring{$\Leftarrow$}{<=})}\label{subsec:soundness}

Here, we show that if $B$ has a clearing sequence, then $G$ has a Hamiltonian
cycle. Fix a clearing sequence. Every flat is destroyed, and only the pyramid
destroys flats. A flat that the pyramid hits is destroyed by that move, so by
Lemma~\ref{lem:onekill}, the pyramid hits every flat other than the blocking
flats exactly once.

\begin{lemma}\label{lem:lines}
As long as the pyramid is on the board, it is neither in a row that initially
contains a flat of $\mathcal{N}$ nor in a column that initially contains a $y$
or $z$ flat. Hence the pyramid hits flats of $\mathcal{N}$ only in vertical moves, and
$y$ and $z$ flats only in horizontal moves.
\end{lemma}

\begin{proof}
We prove the first statement by induction on the number of moves. The starting
cell $(B_\rho,\ D_\rho)$ satisfies it by Lemma~\ref{lem:contents}(1). Now consider a move of the pyramid after
which it is still on the board, and assume that the statement holds before the
move. In this move, the pyramid does not hit the stack.

A horizontal move keeps the pyramid in its row. It ends in column $0$ or $w-1$,
or next to the flat it hits. That flat is in the pyramid's row, so it is not in
$\mathcal{N}$. If it is a $y$ or $z$ flat, the pyramid ends in a column
$\equiv 1$ or $3\pmod 4$; if it is a blocking flat, the pyramid ends in column
$1$ or $2$. None of these columns contains a $y$ or $z$ flat.

A vertical move keeps the pyramid in its column. It ends in row $0$ or $h-1$,
or next to the flat it hits, which is not a $y$ or $z$ flat. If that flat is in
$\mathcal{N}$, the pyramid ends in a row $\equiv 1$ or $3\pmod 4$; if it is a
blocking flat, the pyramid ends in row $1$ or $2$. None of these rows contains a
flat of $\mathcal{N}$.

For the second statement, a flat that the pyramid hits lies on the line along
which it moves, which passes through the cell where the move starts. By the
first statement, this line is a column if the flat is in $\mathcal{N}$, and a
row if the flat is a $y$ or $z$ flat.
\end{proof}

\begin{lemma}\label{lem:exit}
A move of the pyramid that starts outside the departure row $B_i$ can only end in that row by
hitting $v_i^+$ in a downward move. A move that starts outside the arrival row
$A_i$ can only end in that row by hitting $v_i^-$ in an upward move.
\end{lemma}

\begin{proof}
A move that starts outside a row and ends in it is vertical. Neither $B_i$ nor
$A_i$ is row $0$ or $h-1$, so such a move ends either next to the stack, which
destroys the pyramid, or next to the flat it hits, in the row directly above or
below that flat. By Lemma~\ref{lem:contents}(2), rows $B_i+1$ and $A_i-1$ are
empty, row $B_i-1$ contains only $v_i^+$ (nothing if $i=\rho$), and row $A_i+1$
contains only $v_i^-$. So the move can only end in row $B_i$ by hitting $v_i^+$,
which is directly below that row, in a downward move, and it can only end in row
$A_i$ by hitting $v_i^-$, which is directly above that row, in an upward move.
\end{proof}

\noindent Since $v_i^+$ is destroyed when the pyramid hits it, Lemma~\ref{lem:exit} means
that the pyramid enters row $B_i$ at most once, and it starts in that row only if
$i=\rho$. After entering, it stays in row $B_i$ while it moves horizontally,
leaves the row with its first vertical move, and never stops in the row again,
although later moves may still pass through it. We call the part of the play in
which the pyramid is in row $B_i$ its \emph{visit} to row $B_i$. The same holds
for row $A_i$, which the pyramid can only enter by hitting $v_i^-$. By
Lemma~\ref{lem:lines}, the $z$ flats of $v_i$ can only be hit in horizontal
moves along row $B_i$, so both must be hit during the visit to row $B_i$, and
similarly $y(v_i)$ must be hit during the visit to row $A_i$.

\begin{lemma}\label{lem:outer}
Suppose that the pyramid arrives at a cell in row $0$ or $h-1$, or in column $0$
or $w-1$, by moving along a line $L$. Then, until it next hits a flat other
than the blocking flats, it moves only along $L$ and along outer lines.
\end{lemma}

\begin{proof}
By Lemma~\ref{lem:contents}(4), the only flats on outer lines are the blocking
flats. So a move of the pyramid along an outer line ends next to the stack,
which destroys the pyramid, at the end of the line, or next to a blocking flat
that it hits. These cells are a corner or one of $(0,1)$, $(h-1,1)$, $(1,0)$,
$(1,1)$, $(1,w-1)$, $(0,2)$, $(h-1,2)$, $(2,0)$ and $(2,w-1)$, each of which is
an outer crossing, so every later move is again along an outer line.

Now let $s$ be the cell where the pyramid arrives. If $L$ is outer, then by the
above, $s$ is an outer crossing. Otherwise, the other line through $s$ is the
edge of the board, which is outer, and by Lemma~\ref{lem:stack} the stack is not
on $L$. A move back along $L$ then ends at the other end of $L$, which is again
on the edge of the board, or next to a flat that it hits. If that flat is a
blocking flat, the pyramid ends in rows $0$ to $2$ and columns $0$ to $2$, at an
outer crossing.
\end{proof}

\noindent The stack cannot move while both blocking flats are present. To destroy a
blocking flat, the pyramid must stop next to it, at $(0,0)$, $(0,2)$ or $(2,0)$,
since it never reaches $(1,1)$ by the proof of Lemma~\ref{lem:onekill}. Each of
these cells is on the edge of the board and is reached along an outer line, so
by Lemma~\ref{lem:outer} the pyramid never hits another flat other than a
blocking flat afterwards. Hence, in a clearing sequence, the stack does not move
until every flat except at most one blocking flat has been destroyed.

\begin{theorem}\label{thm:soundness}
If $G$ has no Hamiltonian cycle, then $B$ has no clearing sequence.
\end{theorem}

\begin{proof}
Suppose that some sequence of moves clears $B$. As shown above, the pyramid hits
every flat other than the blocking flats exactly once. If some $v_i^+$ with $i\neq\rho$ were not hit in a downward move, then by
Lemma~\ref{lem:exit} the pyramid would never enter row $B_i$, and as it does not
start there, it would never stop in that row; by Lemma~\ref{lem:lines}, the $z$
flats of $v_i$ would then never be hit. So every $v_i^+$ is hit in a downward
move. Likewise, every $v_i^-$ with $i\neq\rho$ is hit in an upward move, since
otherwise $y(v_i)$ is never hit. We now show the following.
\begin{enumerate}
  \item[(a)] After the start (taking $i=\rho$), or after hitting $v_i^+$, the
        next flat of $\mathcal{N}$ that the pyramid hits, if any, is $v_j^-$ or
        $v_k^-$.
  \item[(b)] After hitting $v_i^-$ with $i\neq\rho$, the next flat of
        $\mathcal{N}$ that the pyramid hits, if any, is $v_i^+$.
  \item[(c)] The pyramid hits no flat of $\mathcal{N}$ after $v_\rho^-$.
\end{enumerate}

For (a), the pyramid is at $(B_i,\ D_i)$, at the start of its visit to row
$B_i$. It must hit both $z$ flats of $v_i$ during this visit. They are the only
flats in the row, they lie on opposite sides of the pyramid by \eqref{eq:D}, and
the pyramid cannot pass either of them. So it hits $z(v_i\to v_j)$ in a move to
the left, stopping at $(B_i,\ C_j)$, and $z(v_i\to v_k)$ in a move to the right,
stopping at $(B_i,\ C_k)$. After that, the row is empty, so the pyramid leaves it
by a vertical move along a column $X$ that is $C_j$, $C_k$, $0$ or $w-1$. If
$X=C_\ell$ with $\ell\in\{j,k\}$, then the first move along $X$ either hits
$v_\ell^-$, as required, or ends on the edge of the board, since by
Lemmas~\ref{lem:contents}(3) and~\ref{lem:stack}, $v_\ell^-$ is the only piece
that can be in column $C_\ell$. If $X$ is $0$ or $w-1$, then the pyramid is
already on the edge, having arrived there along row $B_i$. In the remaining
cases, by Lemma~\ref{lem:outer}, the pyramid moves only along $X$, row $B_i$ and
outer lines until it next hits a flat other than the blocking flats. That flat
lies on one of these lines, and since row $B_i$ is now empty, it is $v_j^-$ or
$v_k^-$.

For (b), the pyramid is at $(A_i,\ C_i)$, at the start of its visit to row
$A_i$. It must hit $y(v_i)$, the only flat in this row, during this visit, and
it cannot pass it. By \eqref{eq:side}, $y(v_i)$ lies one cell beyond column
$D_i$ on the side away from column $C_i$, so the pyramid hits it from the side
of column $C_i$ and stops at $(A_i,\ D_i)$. It then leaves the row along one of the columns
$D_i$, $0$ and $w-1$, and as in (a), the next flat of $\mathcal{N}$ that it
hits, if any, is $v_i^+$.

For (c), after hitting $v_\rho^-$, the pyramid is at $(A_\rho,\ C_\rho)$ or
$(A_\rho+2,\ C_\rho)$. By Lemma~\ref{lem:contents}, its row and column now
contain no flats, so its next move takes it along one of them to the edge of the
board. By Lemma~\ref{lem:outer}, it then moves only along that line, $L$, and along
outer lines until it next hits a flat other than the blocking flats. Such a flat
would lie on one of these lines, but $L$ is empty and, by
Lemma~\ref{lem:contents}(4), the outer lines contain only the blocking flats. So
the pyramid only moves along $L$ and the outer lines and never hits a flat of $\mathcal{N}$ again.

By (c), $v_\rho^-$ is the last flat of $\mathcal{N}$ that the pyramid hits.
Since all $2n-1$ flats of $\mathcal{N}$ are hit, every other flat of
$\mathcal{N}$ is followed by another one, so by (a) and (b), the order in which
they are hit is
\[
  v_{\ell_1}^-,\ v_{\ell_1}^+,\ v_{\ell_2}^-,\ v_{\ell_2}^+,\ \ldots,\
  v_{\ell_{n-1}}^-,\ v_{\ell_{n-1}}^+,\ v_\rho^-,
\]
where $v_\rho\to v_{\ell_1}$, each $v_{\ell_t}\to v_{\ell_{t+1}}$, and
$v_{\ell_{n-1}}\to v_\rho$ are arcs of $G$. Each $v_i^-$ is hit exactly once, so
$v_{\ell_1},\ldots,v_{\ell_{n-1}}$ are the $n-1$ vertices other than $v_\rho$,
and $v_\rho,v_{\ell_1},\ldots,v_{\ell_{n-1}},v_\rho$ is a Hamiltonian cycle in
$G$.
\end{proof}

\noindent Subsection~\ref{subsec:completeness} and Theorem~\ref{thm:soundness} show that
$G$ has a Hamiltonian cycle if and only if $B$ has a clearing sequence. Since
$B$ can be built in polynomial time, this proves Theorem~\ref{thm:main}.
\section{No Flats and Equal Pyramids and Stacks}
\label{sec:equal}

In this section, no board contains flats. We call a move \emph{clean} if it destroys exactly one pyramid and one stack. A clean move keeps the numbers of pyramids and stacks equal, so the theorem below follows once we show that a clean move can always be reached.

\equalalwayspossible*

\noindent Recall that a pyramid and a stack are never adjacent before a move: this holds initially, and any such adjacency created by a move is immediately resolved by it. Hence a move destroys pieces only if the moved piece ends up adjacent to pieces of the other kind, and if it ends up adjacent to $j \ge 1$ of them, it destroys those $j$ pieces and itself. A destroying move is therefore clean if and only if the moved piece ends up adjacent to exactly one piece of the other kind. Two pieces in the same row are \emph{consecutive} if all cells between them are empty; two consecutive pieces of different kinds are never adjacent, so at least one empty cell separates them.

\begin{figure}[!ht]
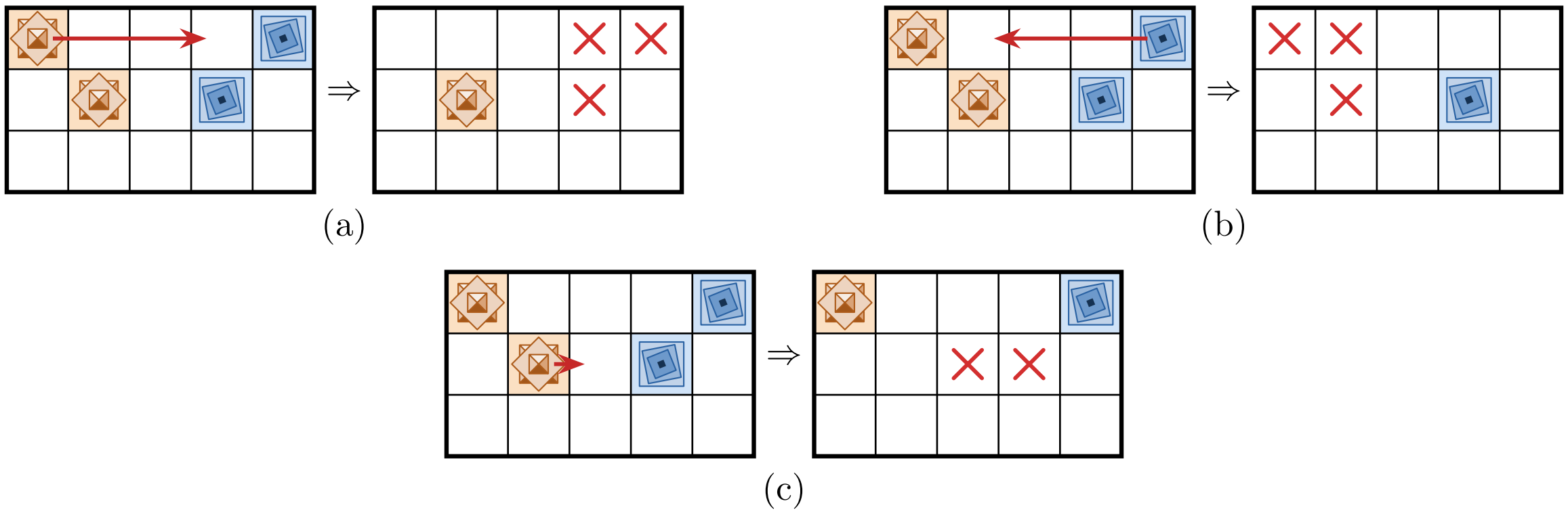

  \centering\setcounter{tlpanel}{0}%
  \setlength\tleqcell{\dimexpr(\linewidth*48/100-2.2em)/10\relax}%
  \ifdim\tleqcell>6mm \setlength\tleqcell{6mm}\fi
  \begin{minipage}[b]{0.48\linewidth}\centering
    \tleqpair{P...K, .P.K., .....}{\draw[tl/move] (tl-0-0) -- (tl-0-3);}{...xx, .P.x., .....}
    \tlpanellabel\label{fig:funnel-a}
  \end{minipage}\hfill
  \begin{minipage}[b]{0.48\linewidth}\centering
    \tleqpair{P...K, .P.K., .....}{\draw[tl/move] (tl-0-4) -- (tl-0-1);}{xx..., .x.K., .....}
    \tlpanellabel\label{fig:funnel-b}
  \end{minipage}

  \medskip
  \begin{minipage}[b]{0.48\linewidth}\centering
    \tleqpair{P...K, .P.K., .....}{\draw[tl/move] (tl-1-1) -- (tl-1-2);}{P...K, ..xx., .....}
    \tlpanellabel\label{fig:funnel-c}
  \end{minipage}
  \caption{Lemma~\ref{lem:funnel} on a board whose top row holds a pyramid and a stack that are consecutive. Moving either of them along the top row is not clean, because each would also destroy the piece below its destination: \ref{fig:funnel-a}~two stacks, \ref{fig:funnel-b}~two pyramids. Between those two pieces, the row below contains consecutive pieces of different kinds, and moving one of them is clean~\ref{fig:funnel-c}.}
  \label{fig:funnel}
\end{figure}

\begin{lemma}\label{lem:funnel}
Let row $r$ contain consecutive pieces of different kinds in columns $a < b$, and suppose that $r$ is the top row or that the cells of row $r+1$ in columns $a+1, \dots, b-1$ are empty. Then some piece in row $r$ or below has a clean move.
\end{lemma}

\begin{proof}
By induction on $b - a \ge 2$. Let the piece $x$ in column $a$ be of kind $\kappa$ and the piece $y$ in column $b$ of kind $\lambda \ne \kappa$, with both in row $r$. If $x$ moves right, it slides to column $b-1$ and comes to rest next to $y$. Its other neighbours are the cell above, which is empty by assumption or off the board, the cell to its left, which is empty or the cell $x$ has just left, and the cell $(r-1, b-1)$ below. So this move is clean unless $(r-1, b-1)$ holds a piece of kind $\lambda$. Symmetrically, moving $y$ to the left is clean unless $(r-1, a+1)$ holds a piece of kind $\kappa$. If $r$ is the bottom row, there is no row below it, so either move is clean. If $b - a = 2$, the two cells below coincide and cannot hold pieces of both kinds, so at least one of the two moves is clean. Otherwise, if neither is clean, row $r-1$ holds a piece of kind $\kappa$ in column $a+1$ and a piece of kind $\lambda$ in column $b-1$, so it contains consecutive pieces of different kinds in some columns $a' < b'$ with $a + 1 \le a' < b' \le b - 1$ (Figure~\ref{fig:funnel}). The cells of row $r$ strictly between columns $a'$ and $b'$ lie strictly between $x$ and $y$ and are empty, and $b' - a' < b - a$, so the induction hypothesis applies to row $r-1$.
\end{proof}

\begin{figure}[t]
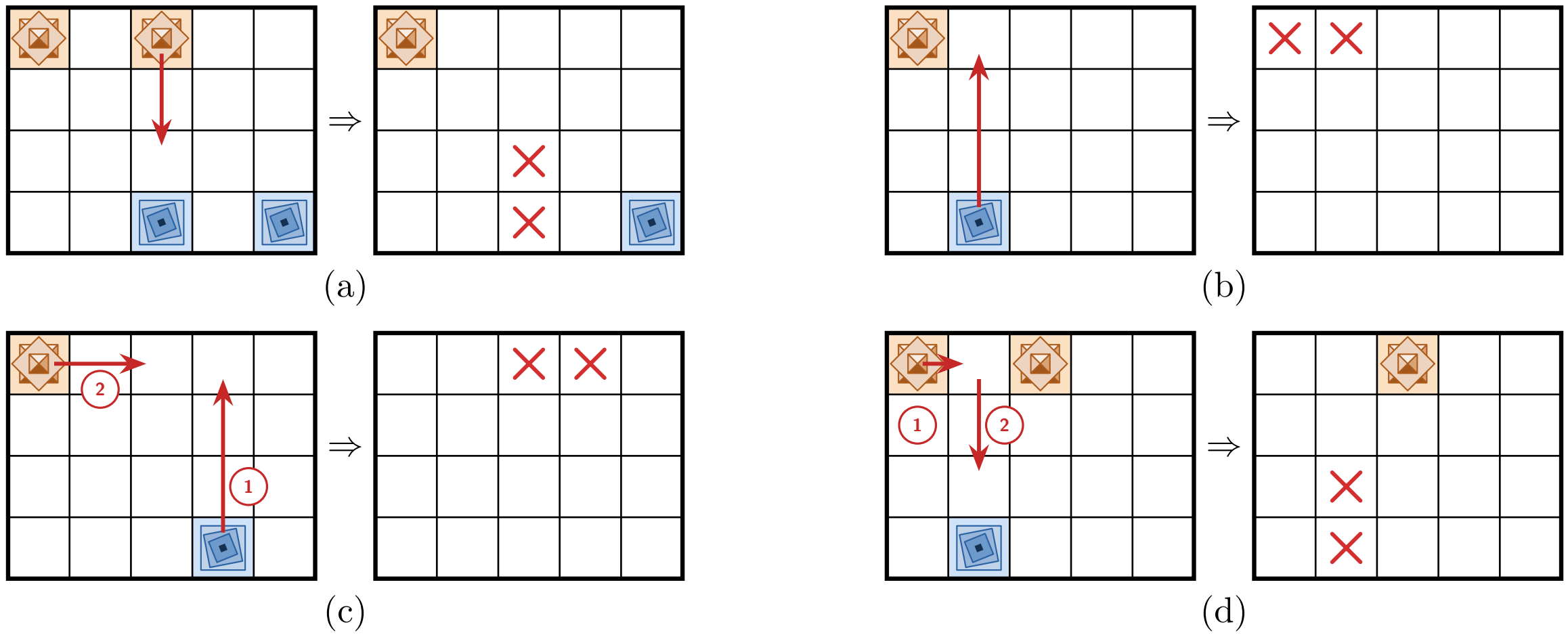

  \centering\setcounter{tlpanel}{0}%
  \setlength\tleqcell{\dimexpr(\linewidth*48/100-2.2em)/10\relax}%
  \ifdim\tleqcell>6mm \setlength\tleqcell{6mm}\fi
  \begin{minipage}[b]{0.48\linewidth}\centering
    \tleqpair{P.P.., ....., ....., ..K.K}{\draw[tl/move] (tl-0-2) -- (tl-2-2);}{P...., ....., ..x.., ..x.K}
    \tlpanellabel\label{fig:top-a}
  \end{minipage}\hfill
  \begin{minipage}[b]{0.48\linewidth}\centering
    \tleqpair{P...., ....., ....., .K...}{\draw[tl/move] (tl-3-1) -- (tl-0-1);}{xx..., ....., ....., .....}
    \tlpanellabel\label{fig:top-b}
  \end{minipage}

  \medskip
  \begin{minipage}[b]{0.48\linewidth}\centering
    \tleqpair{P...., ....., ....., ...K.}{%
      \draw[tl/move] (tl-3-3) -- (tl-0-3); \node[tl/step] at ([xshift=2.5mm]tl-2-3) {1};
      \draw[tl/move] (tl-0-0) -- (tl-0-2); \node[tl/step] at ([yshift=-2.5mm]tl-0-1) {2};}%
      {..xx., ....., ....., .....}
    \tlpanellabel\label{fig:top-c}
  \end{minipage}\hfill
  \begin{minipage}[b]{0.48\linewidth}\centering
    \tleqpair{P.P.., ....., ....., .K...}{%
      \draw[tl/move] ([xshift=-1mm]tl-0-0) -- (tl-0-1); \node[tl/step] at (tl-1-0) {1};
      \draw[tl/move] (tl-0-1) -- (tl-2-1); \node[tl/step] at ([xshift=2.5mm]tl-1-1) {2};}%
      {..P.., ....., .x..., .x...}
    \tlpanellabel\label{fig:top-d}
  \end{minipage}
  \caption{The cases of Lemma~\ref{lem:reach} once the top row holds only pyramids; $z$ is a topmost stack, in the bottom row. \ref{fig:top-a}~A pyramid above $z$ drops onto it. Otherwise $z$ moves up, and the outcome depends on its neighbours in the top row: \ref{fig:top-b}~with one neighbour the move is clean; \ref{fig:top-c}~with none it destroys nothing, and the top row now contains both kinds; \ref{fig:top-d}~with two it would destroy both, so a neighbour first moves over $z$'s column and then drops onto $z$.}
  \label{fig:top}
\end{figure}

\begin{lemma}\label{lem:reach}
Let an $h \times w$ board contain at least one pyramid and at least one stack. Then there is a sequence of at most $w + 2$ moves whose last move is clean and whose other moves destroy nothing.
\end{lemma}

\begin{proof}
Let row $r$ be the topmost row that contains a mover (either a pyramid or a stack). If row $r$ contains both kinds, it contains two consecutive pieces of different kinds, and Lemma~\ref{lem:funnel} gives a clean move. Otherwise, all pieces in row $r$ are of one kind, say $\kappa$, and we write $\lambda$ for the other kind. If $r < h-1$, we move the pieces of row $r$ up one at a time. Each slides to the top row, where its neighbours are empty or of kind $\kappa$, so these at most $w$ moves destroy nothing; we may thus assume that $r$ is the top row. Let $z$ be a topmost piece of kind $\lambda$, in row $r_z < r$ and column $c$. Every piece above row $r_z$ is of kind $\kappa$.

If column $c$ contains a piece above $z$, the lowest such piece slides down and comes to rest directly above $z$. Its other neighbours lie in row $r_z + 1$ or above and are therefore empty or of kind $\kappa$, so the move is clean (Figure~\ref{fig:top}\ref{fig:top-a}). Otherwise, moving $z$ up takes it to the top row, where the cell below it is empty and its two side neighbours are empty or of kind $\kappa$. If exactly one of them is occupied, this move is clean~\ref{fig:top-b}. If neither is, it destroys nothing, and the top row now contains both kinds, so Lemma~\ref{lem:funnel} gives a clean move~\ref{fig:top-c}. If both are occupied, we do not move $z$; instead, the left neighbour moves one cell to the right, into column $c$, where it is blocked by the right neighbour. If $r_z = r - 1$, this move rests next to $z$ and is clean. Otherwise it destroys nothing, and moving the same piece down is clean as before~\ref{fig:top-d}. In every case, at most $w + 2$ moves are used.
\end{proof}

\begin{proof}[Proof of Theorem~\ref{thm:equal}]
We proceed by induction on the number $k$ of pyramids, which equals the number of stacks. If $k = 0$, the board is empty. If $k \ge 1$, Lemma~\ref{lem:reach} yields at most $w + 2$ moves that destroy exactly one pyramid and one stack, after which the board has $k - 1$ pieces of each kind and no flats. By induction, the board can be cleared, using at most $k(w + 2)$ moves in total.
\end{proof}

\noindent The proof is constructive: each step of the strategy can be found in polynomial time, so a clearing sequence of length at most $k(w+2)$ can be computed in polynomial time.

\section{Conclusion and Open Questions}
\label{sec:conclusion}

We have shown that \textsc{Tenebrae Lux} is NP-hard, even on boards with a single pyramid and a single stack, in which case it is NP-complete. On the positive side, every board without flats and with as many pyramids as stacks can be cleared, and a clearing sequence can be computed in polynomial time. Here, we leave several questions open.

First, \textsc{Tenebrae Lux} lies in PSPACE, but our hardness proof uses only one movable piece of each kind. With many movers, the configuration space becomes exponential, as in Atomix, Lunar Lockout and Ricochet Robots, which are all PSPACE-complete~\cite{holzer2004atomix,hartline2003motion,balanza2024ricochet}. Unlike in those games, however, every interaction in \textsc{Tenebrae Lux} destroys pieces, so progress cannot be undone. Is \textsc{Tenebrae Lux} PSPACE-complete? A natural first step is to determine whether some boards require clearing sequences of superpolynomial length, because if none do, then the problem is immediately in NP.

Second, consider boards without flats, with $p$ pyramids and $s$ stacks. Theorem~\ref{thm:equal} only settles the case $p = s$; what if $p \not= s$? A destroying move removes the moved piece and at most three pieces of the other kind, since the cell it has just left is empty, so a board can be cleared only if $s \le 3p$ and $p \le 3s$. The condition $p = s$ cannot simply be relaxed: the board
\begin{center}
  \tlboard[cell=4.5mm]{KK.P, ...., P.K.}
\end{center}
has two pyramids and three stacks, and yet it cannot be cleared, as a short case analysis shows. What is the complexity of \textsc{Tenebrae Lux} on boards without flats? In particular, is it solvable in polynomial time, or does the choice of which pieces destroy each other already make it NP-hard?

Finally, our reduction uses boards with $\Theta(n)$ rows and $\Theta(n^2)$ columns for a graph with $n$ vertices. Does \textsc{Tenebrae Lux} remain NP-hard on boards with a constant number of rows, or with a constant number of flats? Since boards with a constant number of pieces can be solved in polynomial time (Proposition~\ref{prop:membership}), hardness in these settings would have to come from many pieces of some kind.

\paragraph{AI disclosure.} The main insights of the hardness proof, namely using flats to emulate the arcs of a Hamiltonian traversal of a directed graph of bounded out-degree, with a single movable pyramid and a stack trapped in a corner, were developed by the human authors without the use of AI. Later, in discussions of the problem, Anthropic's Claude suggested splitting every vertex into a copy that keeps the incoming arcs and a copy that keeps the outgoing arcs, and helped us choose coordinates that make the construction work. It also observed that boards without flats and with equal numbers of pyramids and stacks always have a clearing sequence, and proposed the proof given in Section~\ref{sec:equal}, which we verified, rewrote and included in the paper. In addition, Claude produced the figures (upon our explicit instructions) and the \LaTeX{} package used to draw the boards, checked the constructions by computer simulation, and helped with proofreading and editing. All AI-generated contributions were verified by the authors, who take full responsibility for the correctness of the results.

\bibliography{bib}

@book{hearn2009games,
  author    = {Hearn, Robert A. and Demaine, Erik D.},
  title     = {Games, Puzzles, and Computation},
  publisher = {A K Peters},
  year      = {2009},
  doi       = {10.1201/b10581}
}

@article{rusu2022hamiltonian,
  author  = {Rusu, Irena},
  title   = {{Hamiltonian} problems in directed graphs with simple
             row patterns},
  journal = {Theoretical Computer Science},
  volume  = {916},
  pages   = {70--85},
  year    = {2022},
  doi     = {10.1016/j.tcs.2022.03.005}
}

@article{savitch1970relationships,
  author  = {Savitch, Walter J.},
  title   = {Relationships between nondeterministic and deterministic
             tape complexities},
  journal = {Journal of Computer and System Sciences},
  volume  = {4},
  number  = {2},
  pages   = {177--192},
  year    = {1970},
  doi     = {10.1016/S0022-0000(70)80006-X}
}

@article{holzer2004atomix,
  author  = {Holzer, Markus and Schwoon, Stefan},
  title   = {Assembling molecules in {ATOMIX} is hard},
  journal = {Theoretical Computer Science},
  volume  = {313},
  number  = {3},
  pages   = {447--462},
  year    = {2004},
  doi     = {10.1016/j.tcs.2002.11.002}
}

@article{hartline2003motion,
  author  = {Hartline, Jeffrey R. K. and Libeskind-Hadas, Ran},
  title   = {The computational complexity of motion planning},
  journal = {SIAM Review},
  volume  = {45},
  number  = {3},
  pages   = {543--557},
  year    = {2003},
  doi     = {10.1137/S003614450139517}
}

@misc{balanza2024ricochet,
  author        = {Balanza-Martinez, Jose and Cantu, Angel A.
                   and Schweller, Robert and Wylie, Tim},
  title         = {A simple proof that {Ricochet Robots}
                   is {PSPACE}-complete},
  year          = {2024},
  doi           = {10.48550/arXiv.2402.11440}
}

@article{flake2002rush,
  author  = {Flake, Gary William and Baum, Eric B.},
  title   = {{Rush Hour} is {PSPACE}-complete, or
             ``{Why} you should generously tip parking lot attendants''},
  journal = {Theoretical Computer Science},
  volume  = {270},
  number  = {1--2},
  pages   = {895--911},
  year    = {2002},
  doi     = {10.1016/S0304-3975(01)00173-6}
}

@article{aloupis2015nintendo,
  author  = {Aloupis, Greg and Demaine, Erik D. and Guo, Alan
             and Viglietta, Giovanni},
  title   = {Classic {Nintendo} games are (computationally) hard},
  journal = {Theoretical Computer Science},
  volume  = {586},
  pages   = {135--160},
  year    = {2015},
  doi     = {10.1016/j.tcs.2015.02.037}
}

@misc{walsh2014candy,
  author        = {Walsh, Toby},
  title         = {{Candy Crush} is {NP}-hard},
  year          = {2014},
  primaryClass  = {cs.CC},
  doi           = {10.48550/arXiv.1403.1911}
}

@inproceedings{guala2014bejeweled,
  author    = {Gual{\`a}, Luciano and Leucci, Stefano and Natale, Emanuele},
  title     = {{Bejeweled}, {Candy Crush} and other match-three games
               are ({NP}-)hard},
  booktitle = {Proceedings of the 2014 IEEE Conference on Computational
               Intelligence and Games (CIG)},
  publisher = {IEEE},
  year      = {2014},
  doi       = {10.1109/CIG.2014.6932866}
}

@incollection{demaine2009playing,
  author    = {Demaine, Erik D. and Hearn, Robert A.},
  title     = {Playing games with algorithms:
               Algorithmic combinatorial game theory},
  booktitle = {Games of No Chance 3},
  editor    = {Albert, Michael H. and Nowakowski, Richard J.},
  series    = {Mathematical Sciences Research Institute Publications},
  volume    = {56},
  pages     = {3--56},
  publisher = {Cambridge University Press},
  year      = {2009},
  doi       = {10.1017/CBO9780511807251.002}
}

@article{kendall2008survey,
  author  = {Kendall, Graham and Parkes, Andrew J. and Spoerer, Kristian},
  title   = {A survey of {NP}-complete puzzles},
  journal = {ICGA Journal},
  volume  = {31},
  number  = {1},
  pages   = {13--34},
  year    = {2008},
  doi     = {10.3233/ICG-2008-31103}
}

@article{uehara2023puzzles,
  author  = {Uehara, Ryuhei},
  title   = {Computational complexity of puzzles and related topics},
  journal = {Interdisciplinary Information Sciences},
  volume  = {29},
  number  = {2},
  pages   = {119--140},
  year    = {2023},
  doi     = {10.4036/iis.2022.R.06}
}

\end{document}